\documentclass[aps,pra,reprint,superscriptaddress]{revtex4-1}

\usepackage{graphicx}
\usepackage{amsmath}
\usepackage{amsthm}
\usepackage{xcolor}

\def\AHU{School of Physics and Optoelectronic Engineering, Anhui University, Hefei, Anhui 230601, China}
\def\PKS{Max-Planck-Institut f\"{u}r Physik komplexer Systeme, D-01187 Dresden, Germany}
\def\ARU{Department of Physics and Astronomy, Aarhus University, DK-8000 Aarhus C, Denmark}

\begin{document}
	\title{Optimal control of quantum state preparation and entanglement creation in two-qubit quantum system with bounded amplitude}

	\author{Xikun Li}
	\affiliation{\AHU}
	\affiliation{\PKS}
	\affiliation{\ARU}

	\begin{abstract}
		We consider the optimal control problem in a two-qubit system with bounded amplitude. Two cases are studied: quantum state preparation and entanglement creation. Cost functions, fidelity and concurrence, are optimized over bang-off controls for different values of the total duration, respectively. For quantum state preparation problem, three critical time points are determined with high precision, and optimal controls are obtained for different durations. A better estimation of the quantum speed limit is obtained, so is the time-optimal control. For entanglement creation problem, two critical time points are determined, one of them is the minimal time to achieve maximal entanglement (unit concurrence) starting from the product state. In addition, the optimal control to reach the unit concurrence is found.
	\end{abstract}

	\maketitle
	
	
	\section{Introduction}
	
	Quantum optimal control (QOC) is crucial to quantum information processing tasks, such as quantum computation and quantum communication. In these tasks complex quantum systems are engineered and manipulated, e.g. to achieve target quantum gates and target quantum states ~\cite{Glaser2015,Alessandro2021,Krotov1993,Brif2010}. In certain cases, the adiabatic operations, which are generally executed very slowly, are desired in experiments, because we wish to avoid heating the sample and to guarantee the target gate/state is prepared with perfect fidelity~\cite{Gericke2007}. However, in experiments the decoherence and noise from the environment often make such slow operations impossible. Therefore, speedup the time evolution by applying fast and robust controls is sensible~\cite{Odelin2019,Chen2010}.
	
	Quantum optimal control theory, which is proposed to solve the problems mentioned above, has been widely applied in various physical systems such as NMR~\cite{KHANEJA2005}, Bose-Einstein condensate~\cite{vanFrank2016}, cold atoms in optical lattices~\cite{Li2018,Srivatsa2021}. One of the important topics in QOC theory is to search the time-optimal control with which the transitions are finished in the \textit{minimal} time. In the context of QOC the minimal time is generally called the quantum speed limit (QSL)~\cite{Caneva2009}. And the temporal shape of the corresponding control field is called time-optimal control. Analytic solutions are available for several cases where the quantum systems considered are in low-dimensional \cite{Lloyd2014,Alessandro2001,Boscain2002,Khaneja2001,Boscain2005,Boscain2006,Boscain2014,Hegerfeldt2013,Hegerfeldt2014,Boozer2012,Jafarizadeh2020}. For multiple-level quantum systems where analytical results are absent, one has to perform numerical optimization.
	
	Roughly speaking, we rely on two classes of optimization: local optimization algorithms, like Krotov~\cite{Sklarz2002}, GRAPE~\cite{KHANEJA2005}, CRAB~\cite{Doria2011}, GROUP~\cite{Sorensen2018} and GOAT~\cite{Machnes2018}, as well as global ones such as differential evolution (DE) and covariance matrix adaptation evolution strategy (CMA-ES)~\cite{Li2018,Zahedinejad2014}. Machine learning techniques, especially reinforcement learning is another promising  method~\cite{Bukov2018a}.
	
	In real experiments the range of tuning parameters of apparatus are finite, thus constraints in general exist on the control field, e.g., the amplitude is bounded. In such cases the appearance of local suboptimal traps in the quantum control landscape makes the QOC problem nontrivial~\cite{Pechen2011,Larocca2018}.
	
	The time-optimal problem of two-qubit system with unbounded amplitude is studied in Ref.~\cite{Jafarizadeh2022}. In this paper we consider the optimal problem in a two-qubit system with bounded amplitude. We consider two problems: quantum state preparation and entanglement creation. For the first one, one wants to achieve the target quantum state with QSL, and to find the temporal shape of time-optimal control. For latter, we are interested in the problem that for given total duration, how large the maximal entanglement can be obtained.
	
	In Ref.~\cite{Li2022} a systematic scenario is proposed to estimate QSL and the time-optimal control by optimizing over the bang-off controls, and the two-level quantum system is considered as an example. Employing the scenario proposed, we optimize over the bang-off controls to estimate the optimal controls. We show in this paper that physics of two-qubit systems is much richer than that of two-level systems.

	\section{Model}\label{sec:Model}

	\begin{figure}
		\includegraphics[width=1\linewidth, trim= 100 230 110 210,clip]{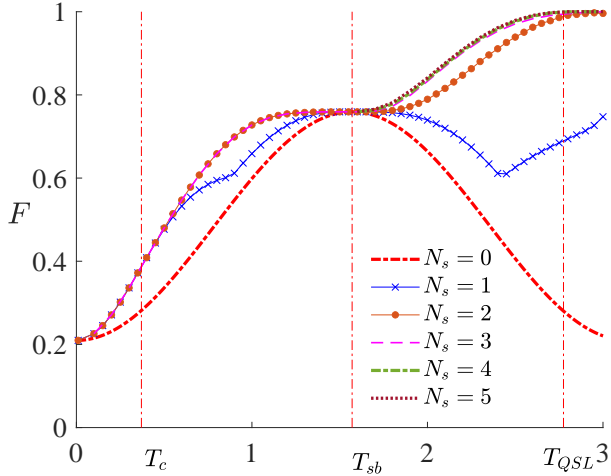}
		\caption{Maximal fidelity $F$ as a function of the total duration $T$ obtained with different number of switches from $N_s=0$ to $N_s=5$. Three critical time points are $T_c = 0.37037$, $T_{\mathrm{sb}}=\pi/2$,
			$T_{\mathrm{QSL}}\approx 2.775$. For $T \in [0,T_c]$, the optimal control field is $P_{T/2}N_{T/2}$. For $T \in (T_c,T_{\mathrm{sb}})$, the optimal control field is $P_{t_1}0_{T-2 t_1}N_{t_1}$. For $T=T_{\mathrm{sb}}$, the optimal control is $0_{\pi/2}$. For $T \in (T_{\mathrm{sb}},T_{\mathrm{QSL}}]$, the best $F$ increases as $N_s$ increases. When $N_s\geq6$, however, the increment of $F$ is too little, thus are not shown in this figure.
		}
		\label{fig:Fid}
	\end{figure}
	
	We consider the symmetrically coupled two-qubit Hamiltonian studied in Ref.~\cite{Bukov2018b}, which is described in the following:
	
	\begin{equation}
		H(t)  =-2g S_1^z S_2^z -h_z(S_1^z+S_2^z)-h_x(t) (S_1^x+S_2^x)
		\label{eq:Hamiltonian}
	\end{equation}
	where $g=h_z=1$ are the interaction strength and static magnetic field along the $z$ direction, and $h_x(t)$ is the time-dependent control field along the $x$ direction. $S_1^z=\sigma_z/2$ is spin-1/2 Pauli operators for the first spin, and so on. The bounded control field $h_x(t)$ is a real function under constraint $|h_x(t)|\leq M$. The dynamics of the system is governed by the Hamiltonian $\mathrm{d} |\psi (t) \rangle / \mathrm{d}t = -\mathrm{i} H(t) |\psi (t) \rangle$, where we set $\hbar=1$, starting from the initial state $|\psi_{i}\rangle$. 
	
	For the quantum state preparation problem, we set the cost function to be the fidelity $F$ defined as follows:
	
	\begin{align}\label{eq:Fidelity}
		F(h_x(t),T) & =|\langle\psi_{t}|\mathcal{T}\exp (- \mathrm{i}\int_0^T H(t) \mathrm{d}t
		|\psi_{i}\rangle|^2 \\ \nonumber
		& =|\langle\psi_{t}|\psi_{f}\rangle|^2.
	\end{align}
	where $\mathcal{T}$ is the time-ordering operator. $T$ is the total duration of time evolution, and $|\psi_{f}\rangle$ is the final state. The initial state $|\psi_{i}\rangle$ is prepared in the ground state of Hamiltonian (\ref{eq:Hamiltonian}) with $h_x=-2$, and the target state $|\psi_{t}\rangle$ is set to be the ground state of Hamiltonian with $h_x=2$.
	
	For the entanglement creation problem, we use concurrence to measure the entanglement of two-qubit pure state. A general two-qubit pure state can be expressed as $| \psi \rangle = a |00\rangle + b |01\rangle + c |10\rangle +
	d |11\rangle$, where $a$, $b$, $c$, $d$ are complex numbers with normalization condition $|a|^2+|b|^2+|c|^2+|d|^2=1$. The concurrence of two-qubit pure state is defined in the following:
	\begin{equation}
		C(| \psi\rangle)  = 2|ad-bc| .
		\label{eq:concurrence}
	\end{equation}
	Specially, we denote $C((h_x(t),T) \equiv C(| \psi_f \rangle)$ the concurrence of final state following the Schr\"{o}dinger evolution with control field $h_x(t)$~\cite{Hill1997}.
	
	For quantum state preparation problem (entanglement creation problem), we want to find the control field $h_x(t)$ which maximizes the fidelity $F(h_x(t),T)$ ($C((h_x(t),T)$) for given $T$. We refer to such $h_x(t)$ as the
	\textit{optimal control} for $T$. Particularly, for quantum state preparation problem, we wish to estimate the quantum speed limit $T_{\mathrm{QSL}}$ with which the target state is obtained with unit fidelity $F=1$. For entanglement creation problem, we want to calculate the minimal time $\tau_{\mathrm{min}}$ such that the unit concurrence $C=1$ is reached. Notice that different from the quantum state preparation problem, the number of two-qubit pure states with unit concurrence is \textit{infinite}, while there is, in general, only \textit{one} target state for quantum state preparation problem.
	
	\begin{figure}
		\includegraphics[width=0.85\linewidth, trim= 80 230 80 230,clip]{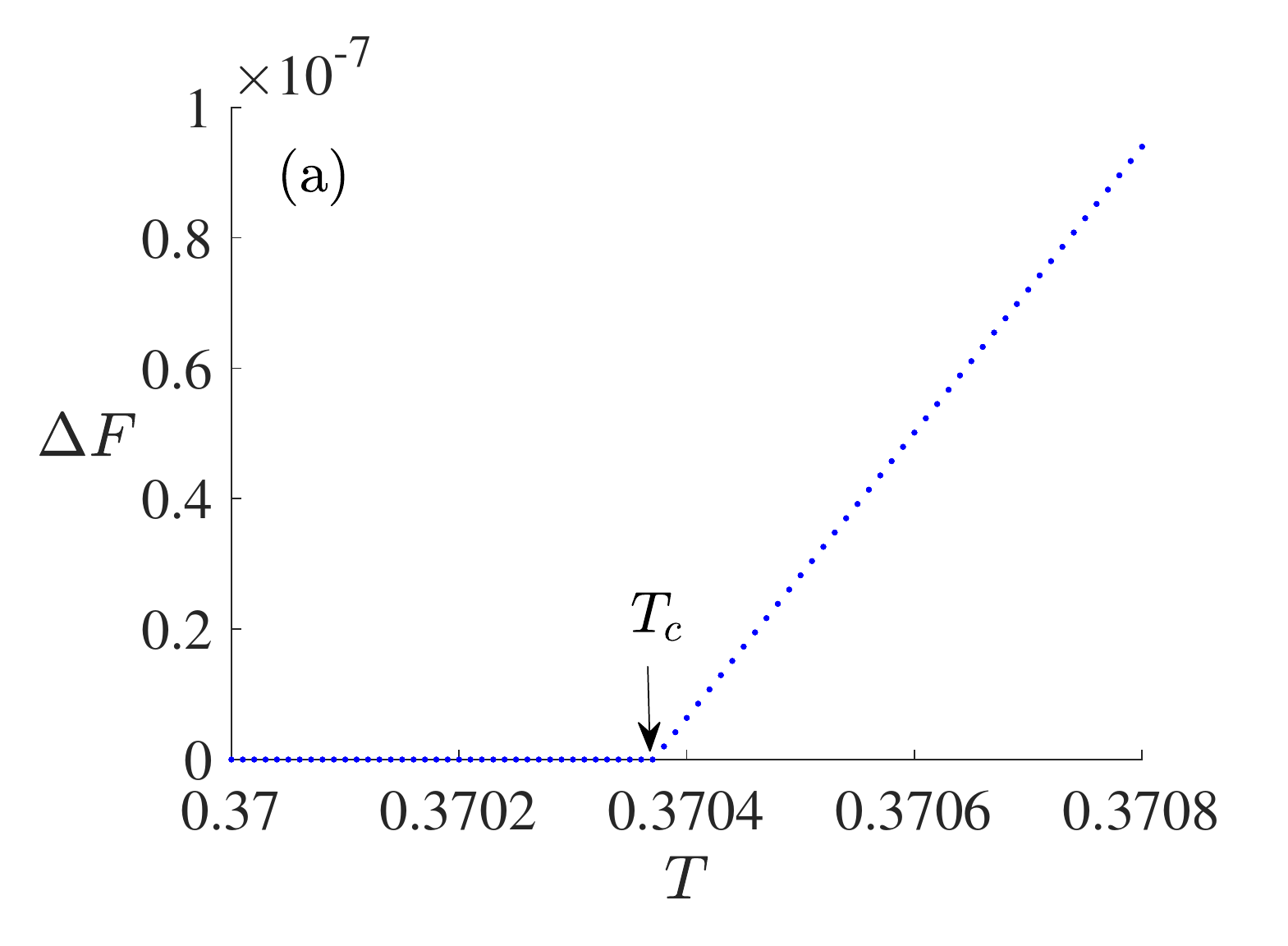}\\
		\includegraphics[width=0.85\linewidth, trim= 80 230 80 230,clip]{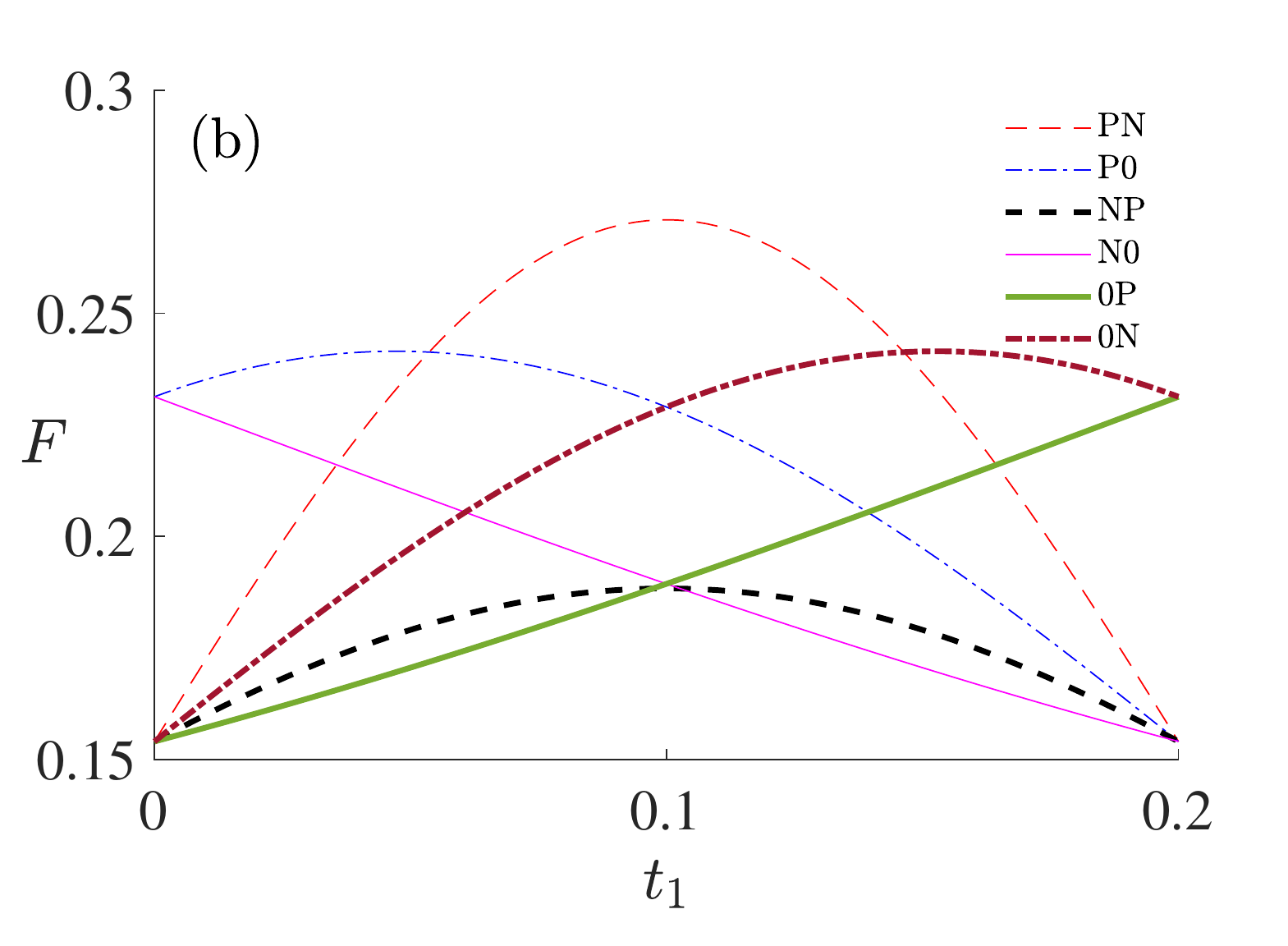}\\
		\includegraphics[width=0.85\linewidth, trim= 80 230 80 230,clip]{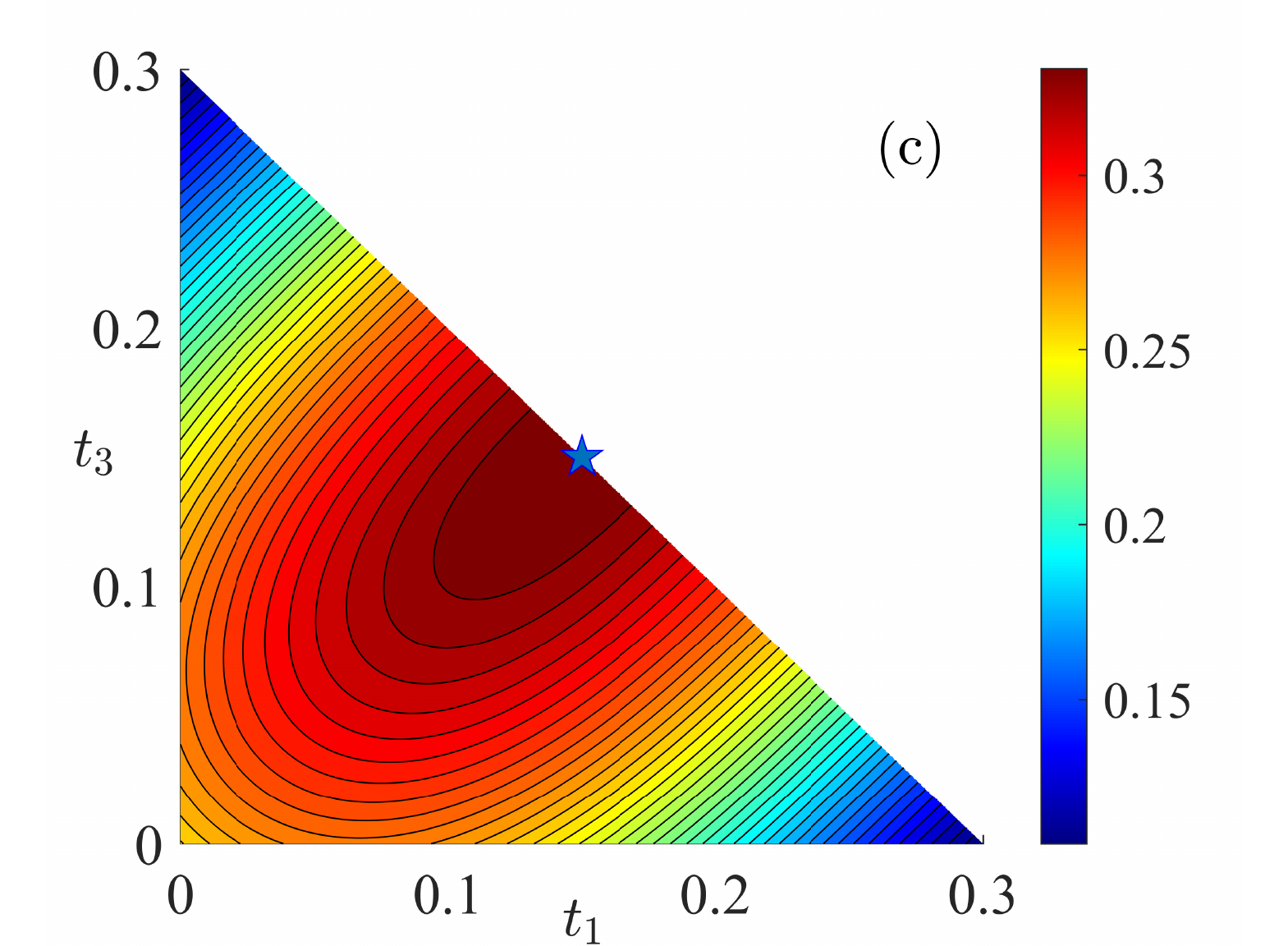}
		\caption{The optimal control is $P_{T/2}N_{T/2}$ for $T \in [0,T_c]$. (a) $\Delta F = F_{2}- F_{1}$ versus the total duration $T$. $F_{N_s}$ is the best fidelity obtained with number of switch $N_s$. $\Delta F=0$ when
			$T\leq T_c$. However, $\Delta F>0$ when $T>T_c$. $T_c=0.37037$ is indicated by an arrow. (b) The fidelity as a function of $t_1$ for all 6 types of $N_s=1$ control with duration vector $\mathbf{t}=[t_1,T-t_1]$ where $T=0.2$. The best fidelity is obtained with control $P_{0.1}N_{0.1}$ (c) The quantum control landscape of fidelity as a function of $[t_1,t_3]$ for $N_s=2$ control $P_{t_1}0_{t_2}N_{t_3}$ with $t_1+t_2+t_3=0.3$. The maximal fidelity is indicated by a blue pentagram whose location is $[t_1,t_3]=[0.15,0.15]$, which means $t_2=0$ and the $N_s=2$ control is reduced to $N_s=1$ control $P_{0.15}N_{0.15}$.
		}
		\label{fig:Tc}
	\end{figure}
	
	We emphasize that Pontryagin's principle does not necessarily mean that the optimal control is always of bang-bang type when the Hamiltonian is linear in the controls. The case of of singular controls exists~\cite{Boscain2006,Bao2018}. The Hamiltonian (\ref{eq:Hamiltonian}) is linear in control $h_x(t)$, and is called bilinear Hamiltonian~\cite{Cong2014}. Therefore, it is sensible to employ the family of \textit{bang-off} controls, (or bang-singular controls), to optimize the cost functions.
	
	Employed the same scenario in Ref~\cite{Li2022}, we optimize $F$ and $C$ over bang-off control. The bang-off control refers to a finite concatenation of \textit{bang} controls $P$ and $N$, and \textit{off} control $0$, which is also called singular control~\cite{Boscain2006}. $P$ ($N$) is short for Positive (Negative) where $h_x(t) = M$ ($h_x(t) = -M$) and $0$ is $h_x(t) = 0$. The control field is represented by the type--a sequence of $P$, $N$ and 0--and vector of durations $\mathbf{t}=[t_1,t_2,...]$. For example, one control field $P_{t_1} 0_{t_2} N_{t_3}$ is defined in the following
	
	\begin{equation}
		h_x(t)= \Bigg\{ \begin{matrix}
			M& 0 \;\;\leq t < t_1 \\
			0 & \;\; t_1\leq t < t_1+t_2 \\
			-M & \;\; t_1+t_2\leq t \leq t_1+t_2+t_3, \\
		\end{matrix}
		\label{eq:bang1}
	\end{equation}
	where the order of letter sequence is from left to right. For the example above, the switch number is two $N_s=2$, and the bang-off control is of type $P0N$ which is switched from bang ($P$) to off ($0$), then to bang ($N$).
	The number of possible types $N_{\mathrm{type}}$ is at most $3\times 2^{N_s}$ for a given number of switches $N_s$. For certain initial/target quantum states, $N_{\mathrm{type}}$ can be further reduced. Here we take $M=4$ such that $|h_x(t) \leq 4|$.
	
	For given $T$, we optimize $F$ (and $C$) starting from $N_s=0$. For each type with given $N_s$ we optimize the vector of durations $\mathbf{t}$ using quasi-Newton method. We denote $F_i$ the maximal fidelity using control fields with $N_s=i$, and the difference of maximal fidelity $\Delta F_{i} \equiv F_{i+1}-F_{i}$. Similar notations are defined for $C$.
	
	Once $\Delta F$ ($\Delta C$) is zero or vanishing small as $N_s$ increases, we stop the optimization and estimate the optimal control with the corresponding optimized control field.
	
	\begin{figure}
		\includegraphics[width=0.85\linewidth, trim= 80 220 70 220,clip]{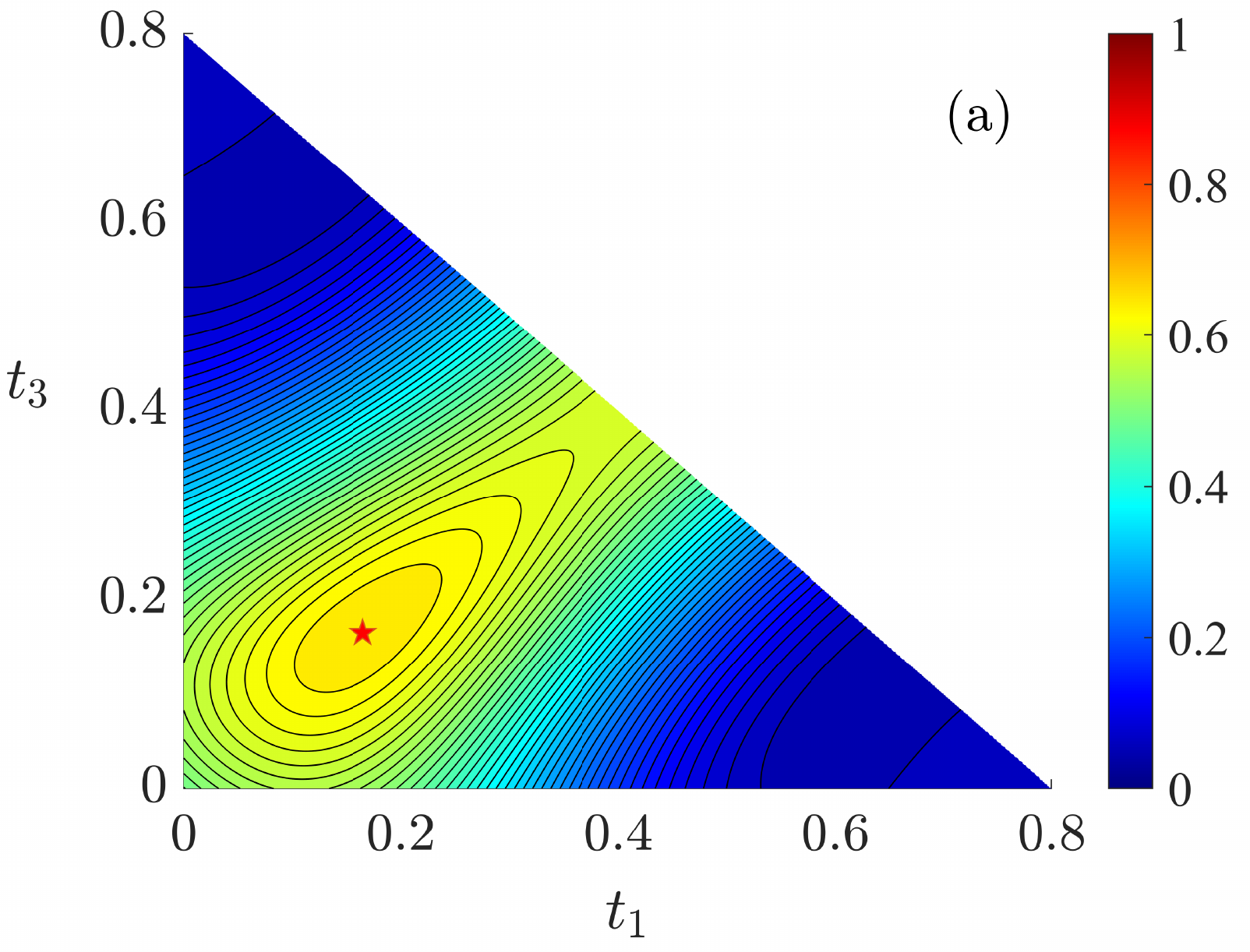}\\
		\includegraphics[width=0.85\linewidth, trim= 80 230 80 230,clip]{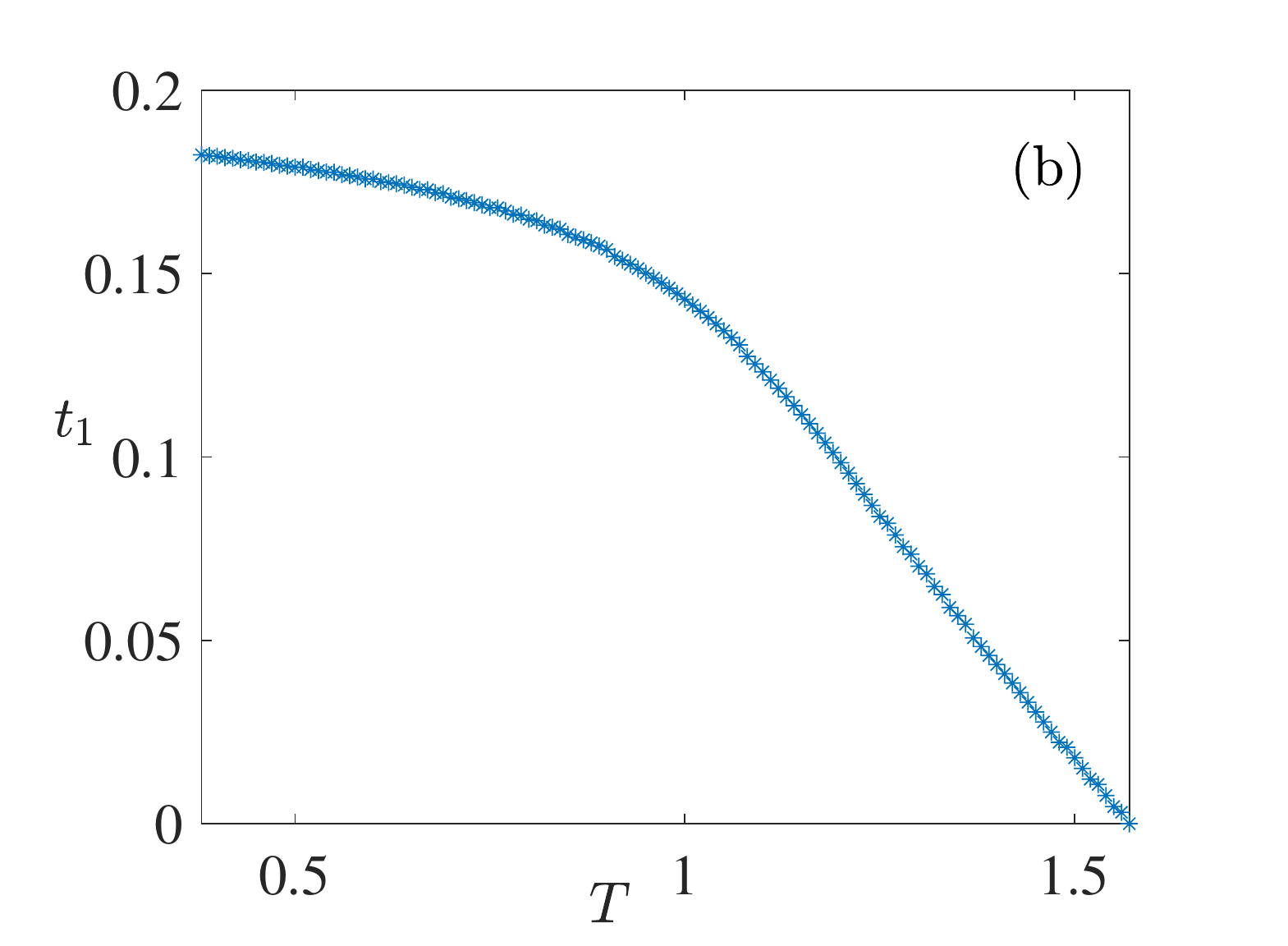}
		\caption{The optimal control is of type $P_{t_1}0_{T-2t_1}N_{t_1}$ for $T \in (T_c,T_{\mathrm{sb}}]$. (a) The quantum control landscape of fidelity as a function of $[t_1,t_3]$ for $N_s=2$ control $P_{t_1}0_{t_2}N_{t_3}$
			with $t_1+t_2+t_3=0.8$. The maximal fidelity is indicated by a red pentagram whose location is $[t_1,t_3]=[0.1648,0.1648]$. (b) $t_1$ as a function of the total duration $T$ for the optimal control  $P_{t_1}0_{T-2t_1}N_{t_1}$. $t_1=0$ when $T=\pi/2$, thus the optimal control is $0_{\pi/2}$.
		}
		\label{fig:Tsb}
	\end{figure}
	
	\begin{figure}
		\includegraphics[width=0.9\linewidth, trim= 70 230 70 230,clip]{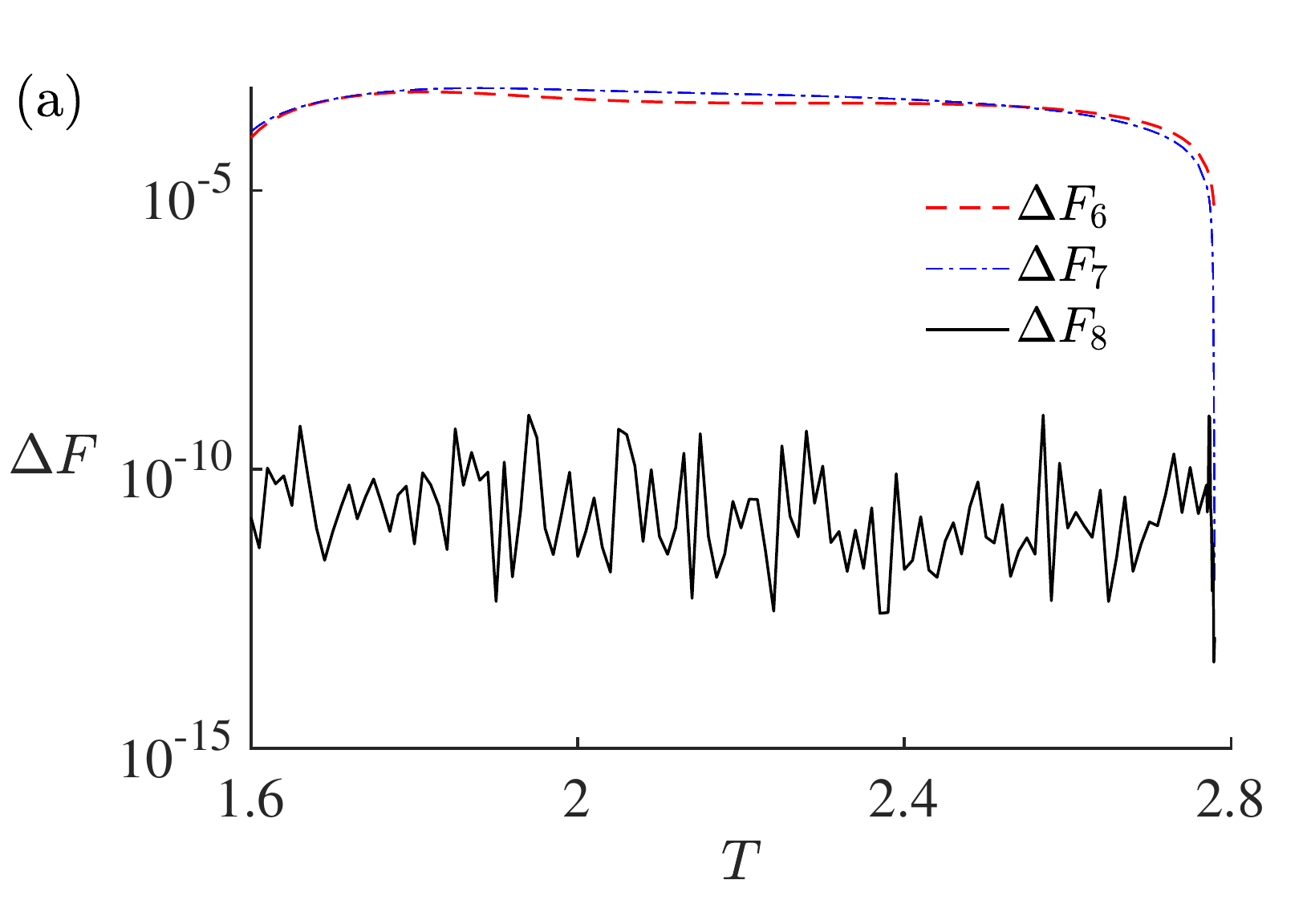}\\
		\includegraphics[width=0.85\linewidth, trim= 100 230 100 230,clip]{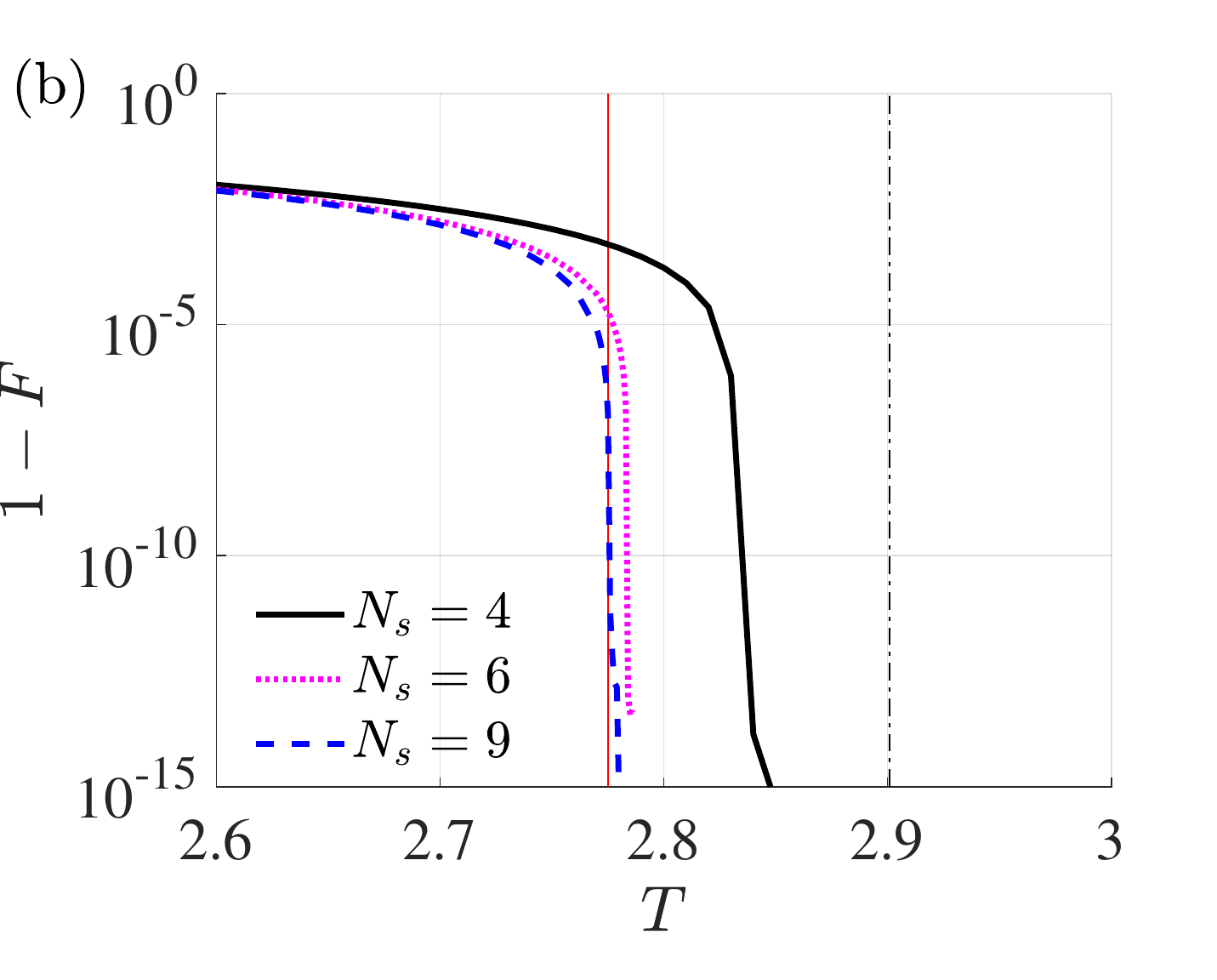}
		\caption{(a) The differences of best fidelity $\Delta F$ obtained with switch number from $N_s=6$ to $N_s=9$ as a function of total duration $T$, where $\Delta F_6 = F_7-F_6$ and so on. (b) The infidelity $1-F$ as a
			function of total duration $T$ with optimal control protocol of $N_s=4$ (black solid line), $N_s=6$ (magenta dotted line), and $N_s=9$ (blue dashed line). The estimation of quantum speed limit is marked by a vertical asymptotic line (blue dashed line), where two vertical lines (red solid line and black dash-dotted line) are the estimation of quantum speed limit obtained by GRAPE $T\approx2.775$ (red solid line) and by symmetric ansatz (black dash-dotted line) $T\approx2.907$ obtained in Ref.~\cite{Bukov2018b}, respectively.
		}
		\label{fig:Tqsl}
	\end{figure}
	
	\begin{figure}
		\includegraphics[width=1\linewidth, trim= 90 230 90 230,clip]{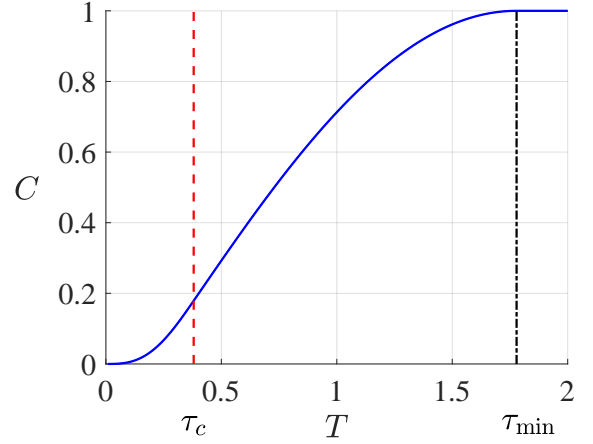}
		\caption{Concurrence C of the optimal control as a function of total duration $T$. Two critical time points are $\tau_c=0.380181$, and $\tau_{\mathrm{min}} \approx 1.778635$.
		}
		\label{fig:concurrence}
	\end{figure}
	
	\begin{figure}
		\includegraphics[width=1\linewidth, trim= 90 230 90 230,clip]{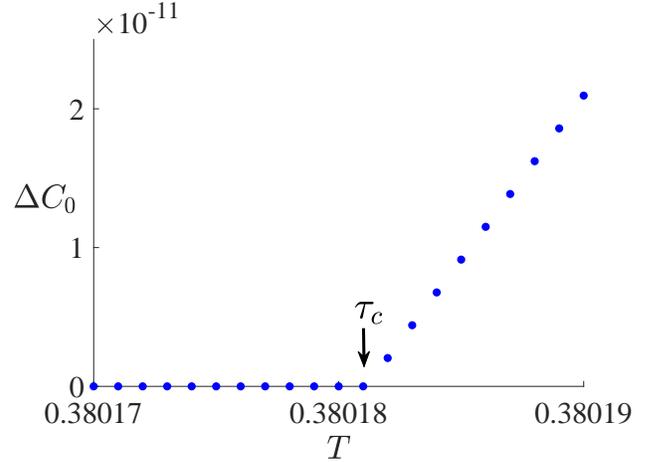}
		\caption{$\Delta C_0 \equiv C_{1}- C_{0}$ versus the total duration $T$. The critical time point is $\tau_c=0.380181$. $\Delta C_0=0$ when $T\leq \tau_c$, and $\Delta C_0 >0$ when $T > \tau_c$. For $T \in [0,\tau_c]$, the optimal control field is $P_T$ and $N_T$.
		}
		\label{fig:tauc}
	\end{figure}
	
	\begin{figure}
		\includegraphics[width=1\linewidth, trim= 90 230 90 230,clip]{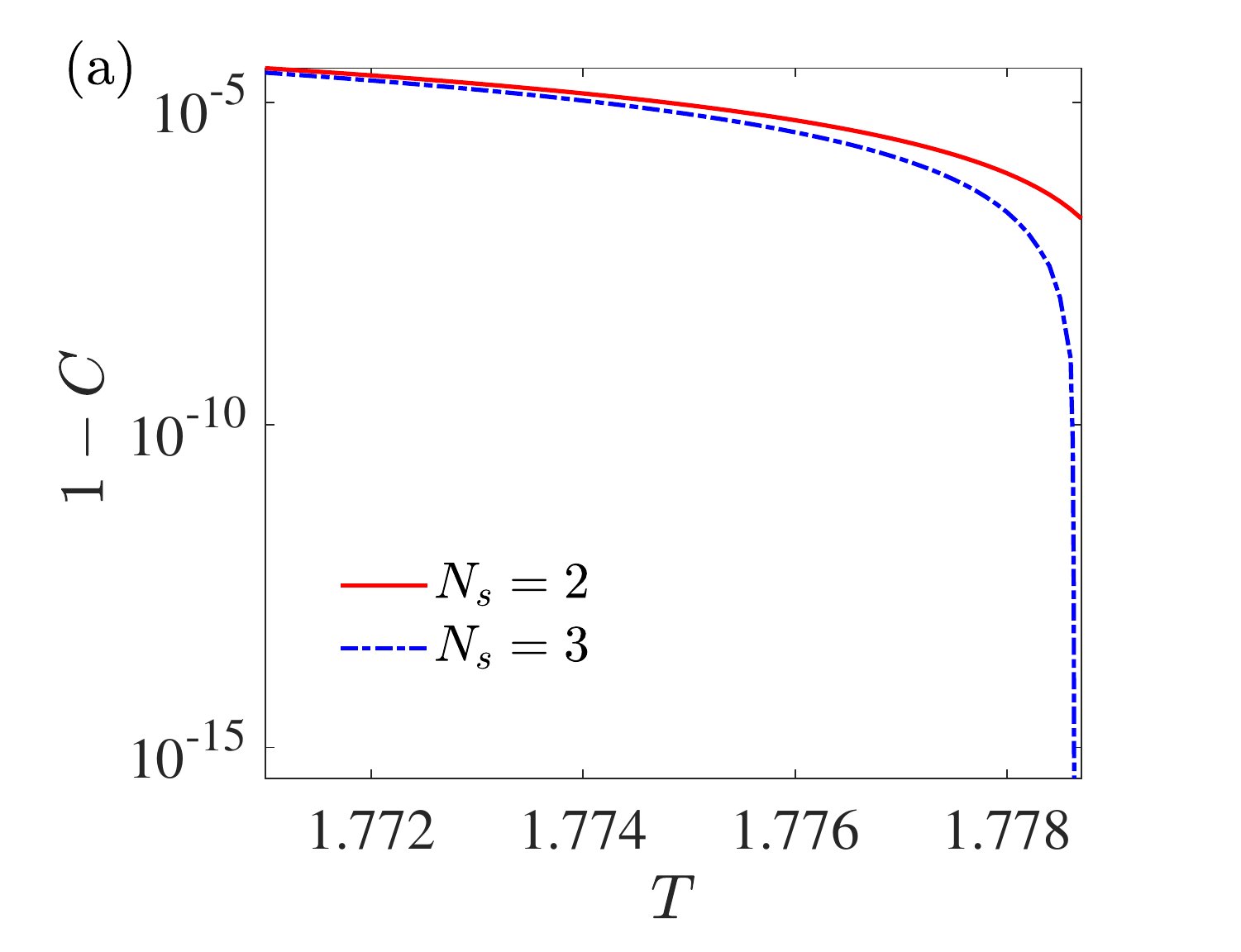}\\
		\includegraphics[width=1\linewidth, trim= 90 230 90 230,clip]{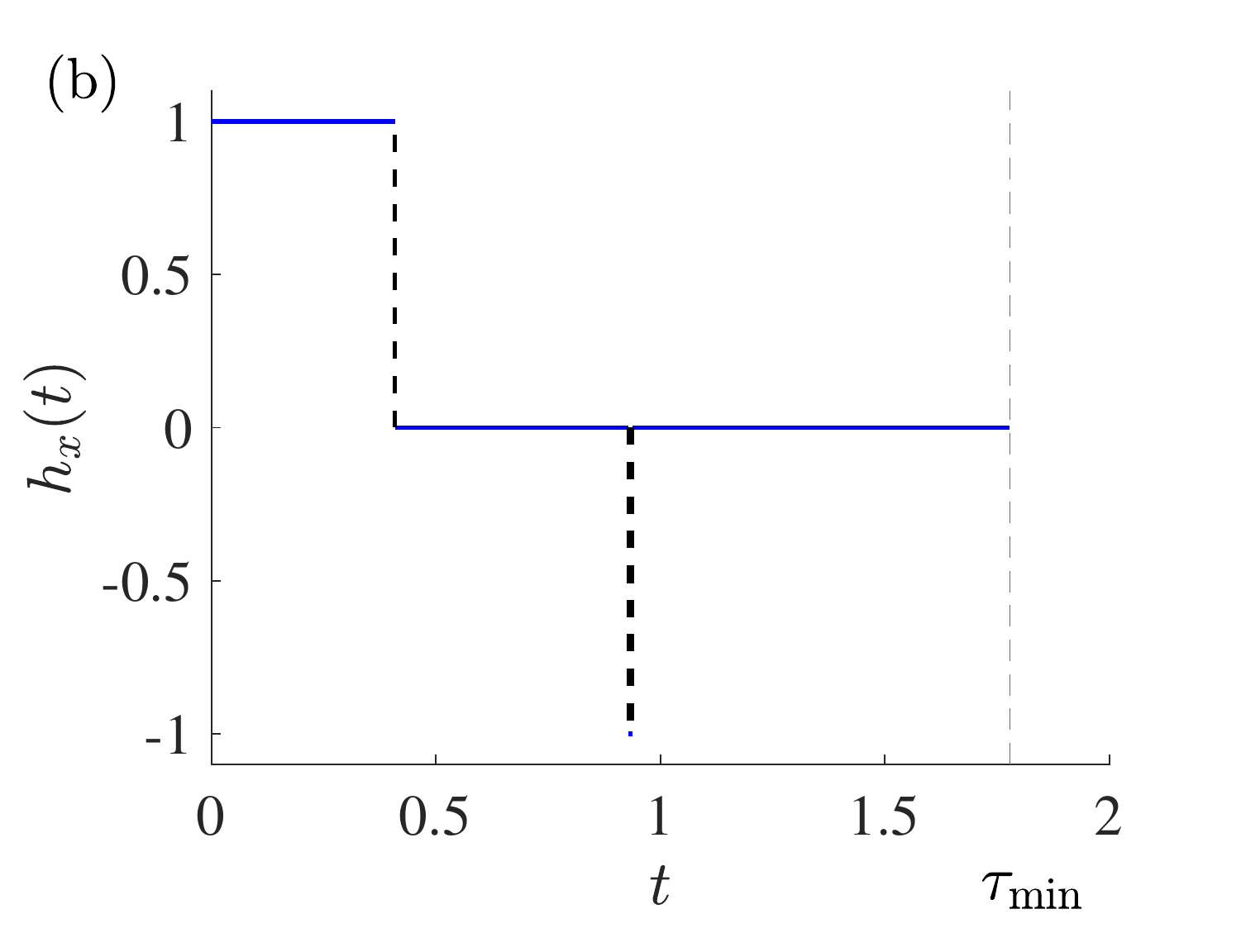}\\
		\caption{(a) Inconcurrence $1-C$ as a function of total duration $T$ for $N_s=2$ and $N_s=3$ with the initial state $|00\rangle$. The unit concurrence is obtained by the optimal control field $P_{t_1}0_{t_2}N_{t_3}0_{t_4}$ which is shown in (b). The estimation of minimal time to reach unit concurrence is $\tau_{\mathrm{min}} \approx 1.778635$.
		}
		\label{fig:taumin}
	\end{figure}
	
	\begin{figure*}
		\includegraphics[width=1\linewidth, trim= 50 580 30 50,clip]{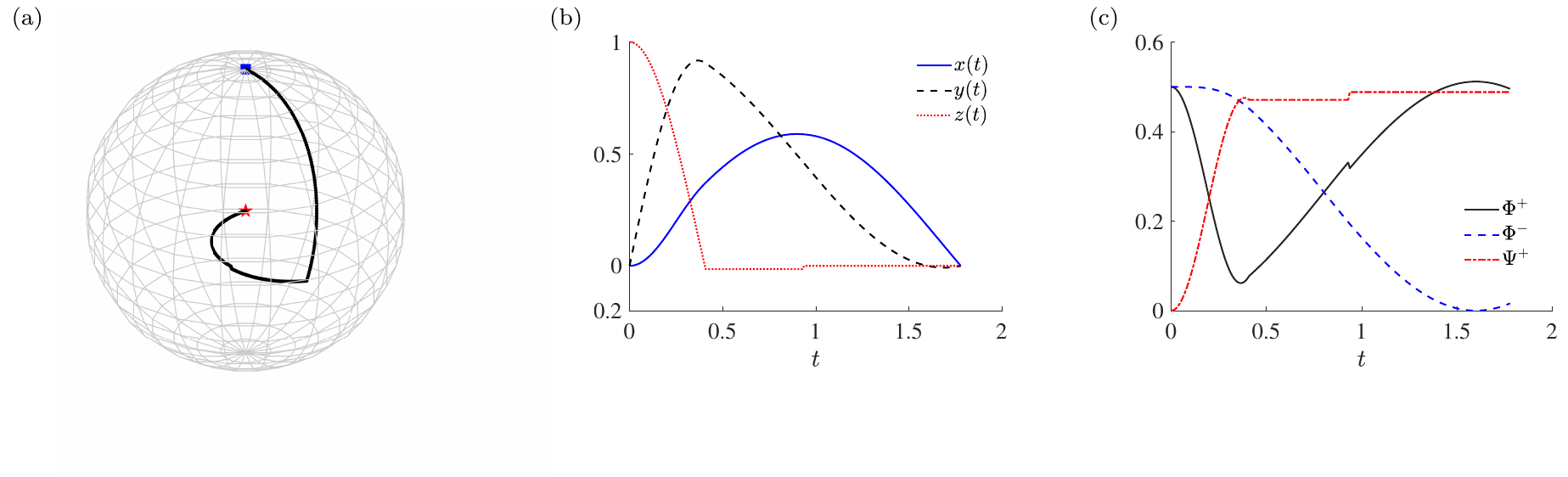}
		\caption{The optimal trajectory obtained using the optimal control shown in Fig.\ref{fig:taumin}(b). (a) The trajectory of reduced density matrix obtained using the optimal control on the Bloch sphere by tracing one qubit. The initial state on the north pole is marked by the blue square, and the final state which is on the centre is marked by a red pentagram. The optimal trajectory is shown in black solid line. (b) The corresponding Cartesian coordinate shown in (a). (c) The coefficients in Eq.~(\ref{eq:coeff}) with respect to three Bell states which span the triplet subspace of two-qubit Hilbert space. 
		}
		\label{fig:bloch}
	\end{figure*}
	
	\section{Quantum State Preparation}\label{sec:state}
	In Ref.~\cite{Bukov2018b} the two-qubit Hamiltonian (\ref{eq:Hamiltonian}) is investigated from the view point of quantum control phase transition. Three critical time points $T_c$, $T_{\mathrm{sb}}$ and $T_{\mathrm{QSL}}$
	are estimated by studying the behavior of several physical quantities, such as correlation. Numerically the bang-bang protocol is optimized using stochastic descent (SD) method to approximate the optimal control fields and to estimate QSL. However, the values of these critical time points are not accurate by optimizing bang-bang control using SD. The optimal controls in different phase regions are approximated by averaging the optimized bang-bang control. However, the optimal controls obtained do not behave like bang-bang control at all.
	
	To overcome the problems mentioned above, we employ the bang-off controls instead. We obtain the values of three critical time points by observing the behavior of $\Delta F_i$. In addition, we find that the optimal controls in overconstrained phase ($T \in (0,T_c]$) and correlated and glassy phase ($T \in (T_c,T_{\mathrm{sb}}]$) are indeed bang-off controls. The values of critical time points are more accurate than those obtained using bang-bang control~\cite{Bukov2018b}. They are $T_c=0.37037$, $T_{\mathrm{sb}}=\pi/2$, and $T_{\mathrm{QSL}}\approx 2.775$; cf.Fig.~\ref{fig:Fid}. In addition, we find that the optimal types found with different $N_s$ is of type $P...N$. This might result from the fact that the initial sate is the ground state of Hamiltonian with negative value $h_x=-2$, while the target state is that of positive value $h_x=+2$. The details are in the following.
	
	\subsection{Time-optimal control for $T \in (0,T_c]$}\label{subsec:Tc}
	
	For $T \in (0,T_c]$, the optimal control protocol is $P_{T/2}N_{T/2}$ in the overconstrained region. The best fidelity obtained with larger $N_s$ is equal to that with $P_{T/2}N_{T/2}$ for $T \in (0,T_c]$. In addition, the optimal control fields with larger $N_s$ reduce to $P_{T/2}N_{T/2}$.
	
	In Fig.~\ref{fig:Tc}(a) we show $\Delta F=F_2-F_1$, which is the difference between the best fidelity obtained with $N_s=2$ and that with $N_s=1$. It is observed that $\Delta F=0$ for $T \leq 0.37037$, whereas $\Delta F>0$ for $T > 0.37037$. The same result holds for $N_s\geq 3$. In such way we locate the value of $T_c=0.37037$ which is more accurate than the one obtained in \cite{Bukov2018b} where $T_c$ is approximately equal to 0.38.
	
	In Fig.~\ref{fig:Tc}(b) we show that the maximal fidelity is obtained with control $P_{T/2}N_{T/2}$ for $T=0.2$, within all six types of bang-off control with $N_s=1$. The same conclusion is true for all $T \in (0,T_c]$. In Fig.~\ref{fig:Tc}(c) we further demonstrate that the optimal control is $P_{0.15}0_{0}N_{0.15}$, which is one of twelve types of $N_s=3$ control, for $T=0.3$. Notice that $P_{0.15}0_{0}N_{0.15}$ is in fact  $P_{0.15}N_{0.15}$, thus belongs to the type $P_{T/2}N_{T/2}$. Same result holds for $N_s \geq 4$.
	
	Therefore, we have numerically demonstrate that the optimal control is  $P_{T/2}N_{T/2}$ for $T \in (0,T_c]$ with $T_c=0.37037$.
	
	\subsection{Time-optimal control for $T \in (T_c,T_{\mathrm{sb}}]$}
	
	For $T \in (T_c,T_{\mathrm{sb}}]$, the optimal control protocol is of $N_s=2$ type $P_{t_1}0_{t_2}N_{t_1}$ with $2 t_1 + t_2=T$. In addition, the optimal control fields with $N_s \geq 3$ reduce to $P_{t_1}0_{t_2}N_{t_1}$.
	
	For $N_s=2$, we have numerically checked that the optimal control is of type $P0N$ within all 12 types when $T \in (T_c,T_{\mathrm{sb}}]$. Moreover, the best fidelity is obtained with a special control field
	$P_{t_1}0_{T-2t_1}N_{t_1}$, i.e., the first duration being equal to the last one $t_3=t_1$. In Fig.~\ref{fig:Tsb}(a) it is shown that the quantum control landscape of control field $P_{t_1}0_{T-t_1-t_3}N_{t_3}$ with $T=0.8$. The maximal fidelity is obtained with $t_1=t_3=0.1648$. Similar results hold for $T \in (T_c,T_{\mathrm{sb}}]$. Therefore, the optimal duration vector is $[t_1,T-2t_1,t_1]$. The value of $t_1$, shown in Fig.~\ref{fig:Tsb}(b), is determined numerically. Notice that $t_1=0$ when $T=T_{\mathrm{sb}}$, thus the optimal control field reduces to $0_{\pi/2}$.
	
	In Ref.~\cite{Bukov2018b} the temporal shape of optimal control for $T  \leq T_{\mathrm{sb}}$ is obtained by averaging the optimized bang-bang controls, which turns out to be approximately bang-off type. Considering the above results and the ones obtained here, we conjecture that the optimal controls for $T \in (0,T_{\mathrm{sb}}]$ are bang-off controls.
	
	\subsection{Time-optimal control for $T \in (T_{\mathrm{sb}}, T_{\mathrm{QSL}}]$}
	
	In the symmetry-broken phase region $T \in (T_{\mathrm{sb}}, T_{\mathrm{QSL}}]$, the double degeneracy of optimal control field is displayed by two optimal control fields of same type, but with duration vectors where one is the flipped vector of another, i.e. $h_1^{\mathrm{opt}}(t)=-h_2^{\mathrm{opt}}(T-t)$.
	
	For $T\in (T_{\mathrm{sb}}, T_{\mathrm{QSL}})$, in general, the best fidelity obtained increases as $N_s$ increases. See Fig.~\ref{fig:Fid}. When $N_s \geq 4$, the increment is smaller than $10^{-4}$. In
	Fig.~\ref{fig:Tqsl}(a) we show $\Delta F$ as a function of $T$ for switch numbers from $N_s=6$ to $N_s=9$. We observe that while the difference $\Delta F_6$ and $\Delta F_7$ is of  order $10^{-5}$, $\Delta F_8 \lesssim 10^{-10}$ is vanishingly small. Therefore, we estimate the optimal control in the symmetry-broken phase region by using bang-off control with $N_s=9$.
	
	In Ref.~\cite{Bukov2018b} the quantum speed limit is estimated by optimizing the three-pulse symmetric anstaz, which is in fact bang-off control of one special type $P_{t_1}0_{t_2}P_{t_3}0_{t_4}N_{t_3}0_{t_2}N_{t_1}$ with $N_s=6$. The estimation of quantum speed limit obtained using this anstaz is $T \approx 2.907$, while another estimation using GRAPE is $T_{\mathrm{GRAPE}} \approx 2.775$. In fact, the estimation of $T_{\mathrm{QSL}}$ can be better if the bang-off control is not restricted to be symmetric. From Fig.~\ref{fig:Tqsl}(b) we observe that the estimation of $T_{\mathrm{QSL}}$ using $N_s=4$ bang-off controls is better than the anstaz.    
	
	In Fig.~\ref{fig:Tqsl}(b) we show the infidelity $1-F$ as a function of $T$ obtained using the bang-off control. The optimal control protocol with $N_s=9$ is of type $P0N0NP0P0N$. The unit fidelity is reached
	$F=1-\mathcal{O} (10^{-15})$ with two optimal duration vectors: $\mathbf{t}^*= [0.232,0.244,0.561,0.317,0.017,0.093,0.858,0.044,0.241,\newline 0.173]$, and another one which is the flipped vector of $\mathbf{t}^*$. The estimation of QSL $T_{\mathrm{QSL}}\approx 2.775$ by using bang-off control with $N_s=9$ is indicated by the vertical asymptote which is equal to the one obtained with GRAPE, and less than the one obtained using bang-bang controls or variational ansatz which in fact belongs to one type of $N_s=6$ bang-off controls~\cite{Bukov2018b}. While the control fields obtained using GRAPE are continuous, the temporal shape of bang-off control is much simpler than the former.
	
	\section{Entanglement creation}\label{sec:entanglement}
	
	In this section we investigate the optimal control problem of entanglement creation. For given $T$, we maximize $C$ starting from the initial state $|00\rangle$ which is a product state with $C=0$. Similar to the quantum state preparation problem, two critical time points are found: $\tau_c=0.380181$, and $\tau_{\mathrm{min}} \approx 1.778635$. See Fig.~\ref{fig:concurrence} for illustration.

	\subsection{Time-optimal control for $T \in [0,\tau_c]$}

	For $T \in [0,\tau_c]$ the optimal control is $P_T$ and $N_T$. This is numerically confirmed by two steps. First, we calculate $\Delta C_i \equiv C_{i+1} - C_i$, where $C_i$ is the maximal concurrence obtained with $N_s=i$, for various values of switch number. All these $\Delta C_i$ are zero when $T \leq 0.380181$. However, $C_1 > C_0$ when $T > 0.380181$.  Second, when $T \in [0,\tau_c]$,  we find that $C_0$ is obtained with the control field $P_T$ and $N_T$, and $C_i$'s ($i \geq 1$) are obtained with $P_T$ and $N_T$, too. Therefore, the first critical time point is $\tau_c =0.380181$, and the optimal control is $P_T$ for $T \in [0,\tau_c]$.
	
	In Fig.~\ref{fig:tauc} we show $\Delta C_0$ as a function of $T$ for example. Following the terminology in Ref.~\cite{Bukov2018b}, we call the region with $T \in [0,\tau_c]$ the overconstrained phase, because the optimization is easy, even though the number of global optima is double, rather than single.

	\subsection{Time-optimal control for $T \in (\tau_c, \tau_{\mathrm{min}}]$ }
	
	We have numerically calculate the concurrence $C$ with different values of $N_s$ for $T>\tau_c$. We find that $C$ stops increasing when $N_s$ is larger than three. In addition, the optimized time vectors found by different $N_s$ ($N_s \geq 3$) are very close to each other. Therefore, we conjecture that the optimal control belongs to the control field with $N_s=3$ for $T>\tau_c$.
	
	In Fig.~\ref{fig:taumin}(a) we show the inconcurrence $1-C$ as a function of $T$ for number of switches $N_s=2$ and $N_s=3$. The data for $N_s\geq 4$ is not shown, because the values obtained with $N_s\geq 4$ are equal to those with $N_s=3$. The unit concurrence is obtained with the optimal control when $T=1.778635$, while for $T<1.778635$ the unit concurrence cannot be reached. See Fig.~\ref{fig:taumin}(a) for illustration. Therefore, the minimal time to reach the unit concurrence is estimated to be $\tau_{\mathrm{min}} \approx 1.778635$. The corresponding optimal control estimated with $N_s=3$ control field is $P_{t_1}0_{t_2}N_{t_3}0_{t_4}$ with the optimal time vector being $\mathbf{t}=[0.40858, 0.52057, 8.1384\times 10^{-3},  0.84135]$; cf. Fig.~\ref{fig:taumin}(b). 
	
	In Fig.~\ref{fig:bloch}(a) we show the trajectory of reduced density matrix on the Bloch sphere by tracing one qubit. The initial state is indicated by the blue square, and the final state in the centre of Bloch sphere is indicated by a red pentagram, which means the final state of two-qubit state is one of maximally entangled two-qubit state, i.e. $C=1$. In Fig.~\ref{fig:bloch}(b) we further show the Cartesian coordinate $[x(t),y(t),z(t)]$ of the optimal trajectory on the Bloch sphere. 
	
	We have studied the time evolution of optimal trajectory in a single-qubit picture. Now we investigate it in the full two-qubit picture. Notice that the initial state $|00\rangle$ is inside the Hilbert subspace of triplet manifold, and that the Hamiltonian (\ref{eq:Hamiltonian}) is invariant by exchanging the two qubits~\cite{Bukov2018b}. Therefore, the time evolution of two-qubit system is inside the Hilbert subspace of triplet manifold. The time evolving state is the superposition of three Bell states, i.e., $|\Phi^+\rangle=(|00\rangle+|11\rangle)/\sqrt{2}$,$|\Phi^-\rangle=(|00\rangle-|11\rangle)/\sqrt{2}$, and $|\Psi^+\rangle =(|01\rangle+|10\rangle)/\sqrt{2}$. We monitor the three coefficients with respect three Bell states 
	\begin{equation}
		|\psi(t)\rangle = C_1(t)|\Phi^+\rangle + C_2(t)|\Phi^-\rangle + C_3(t)|\Psi^+\rangle.
		\label{eq:coeff}
	\end{equation}   
	In Fig.~\ref{fig:bloch}(c) we show the squared coefficients $|C_i(t)|^2$ of the quantum state following the optimal trajectory. From Fig.~\ref{fig:bloch}(c) we show that the final state with unit concurrence is not Bell state, because all coefficients are not zero.

	\section{Conclusions}\label{sec:conclusions}
	
	In this paper we investigate the optimal control problem in a symmetrically coupled two-qubit system with bounded amplitude. By optimizing over the family of bang-off controls, the problems of quantum state preparation and entanglement creation are studied. Given the initial states, the cost functions, fidelity and concurrence, are optimized for various durations for the problems mentioned above. By studying the difference of best cost function obtained with different types of control field, optimal controls and critical time points are determined more accurately than the previous work.
	
	For the quantum state preparation problem, we have shown that for durations before symmetry-broken phase occurs, the optimal control fields are indeed of type bang-off, which is taken as an ansatz in the previous work. For durations in the symmetry broken phase, we estimate the optimal control using bang-off controls up to $N_s=9$. As the number of switch increases, the increment of best fidelity is decreasing and approaching vanishingly small. Therefore, we estimate the optimal control field in the symmetry-broken phase with bang-off controls with $N_s=9$. Furthermore, we estimate the quantum speed limit and time-optimal control field using the same bang-off controls. The QSL obtained is equal to that obtained with GRAPE, but the temporal shape of time-optimal control field is simple.
	
	For the entanglement creation problem, we start from the product state and maximize the concurrence using bang-off controls for different durations. Two critical time points are obtained. For durations in the overconstrained phase, the optimal control is simple: the control field takes either the maximal value or the minimal. As the duration increases, the optimal control is of bang-off type with $N_s=3$. The minimal duration to reach the unit concurrence is estimated, and time-optimal control is obtained. The optimal trajectory is also shown using different methods, e.g., the trajectory of the reduced density matrix on a Bloch sphere, and the coefficients of time-evolving with respect the basis of triplet states.   
	
	In the previous work it has been proved that the time-optimal control with a bilinear Hamiltonian in a two-level quantum system with bounded amplitude is indeed bang-off control. Considering the results in the previous work and the ones in this paper, it is interesting to ask whether the same conclusion holds for the two-qubit system with bounded amplitude. An analytical proof is desired.

	\begin{acknowledgments}
		We are grateful to Marin Bukov and Anne Nielsen for helpful discussions.
	\end{acknowledgments}


\begin{thebibliography}{37}%
		\makeatletter
		\providecommand \@ifxundefined [1]{%
			\@ifx{#1\undefined}
		}%
		\providecommand \@ifnum [1]{%
			\ifnum #1\expandafter \@firstoftwo
			\else \expandafter \@secondoftwo
			\fi
		}%
		\providecommand \@ifx [1]{%
			\ifx #1\expandafter \@firstoftwo
			\else \expandafter \@secondoftwo
			\fi
		}%
		\providecommand \natexlab [1]{#1}%
		\providecommand \enquote  [1]{``#1''}%
		\providecommand \bibnamefont  [1]{#1}%
		\providecommand \bibfnamefont [1]{#1}%
		\providecommand \citenamefont [1]{#1}%
		\providecommand \href@noop [0]{\@secondoftwo}%
		\providecommand \href [0]{\begingroup \@sanitize@url \@href}%
		\providecommand \@href[1]{\@@startlink{#1}\@@href}%
		\providecommand \@@href[1]{\endgroup#1\@@endlink}%
		\providecommand \@sanitize@url [0]{\catcode `\\12\catcode `\$12\catcode
			`\&12\catcode `\#12\catcode `\^12\catcode `\_12\catcode `\%12\relax}%
		\providecommand \@@startlink[1]{}%
		\providecommand \@@endlink[0]{}%
		\providecommand \url  [0]{\begingroup\@sanitize@url \@url }%
		\providecommand \@url [1]{\endgroup\@href {#1}{\urlprefix }}%
		\providecommand \urlprefix  [0]{URL }%
		\providecommand \Eprint [0]{\href }%
		\providecommand \doibase [0]{http://dx.doi.org/}%
		\providecommand \selectlanguage [0]{\@gobble}%
		\providecommand \bibinfo  [0]{\@secondoftwo}%
		\providecommand \bibfield  [0]{\@secondoftwo}%
		\providecommand \translation [1]{[#1]}%
		\providecommand \BibitemOpen [0]{}%
		\providecommand \bibitemStop [0]{}%
		\providecommand \bibitemNoStop [0]{.\EOS\space}%
		\providecommand \EOS [0]{\spacefactor3000\relax}%
		\providecommand \BibitemShut  [1]{\csname bibitem#1\endcsname}%
		\let\auto@bib@innerbib\@empty
		\bibitem [{\citenamefont {Glaser}\ \emph {et~al.}(2015)\citenamefont {Glaser},
			\citenamefont {Boscain}, \citenamefont {Calarco}, \citenamefont {Koch},
			\citenamefont {K{\"o}ckenberger}, \citenamefont {Kosloff}, \citenamefont
			{Kuprov}, \citenamefont {Luy}, \citenamefont {Schirmer}, \citenamefont
			{Schulte-Herbr{\"u}ggen}, \citenamefont {Sugny},\ and\ \citenamefont
			{Wilhelm}}]{Glaser2015}%
		\BibitemOpen
		\bibfield  {author} {\bibinfo {author} {\bibfnamefont {S.~J.}\ \bibnamefont
				{Glaser}}, \bibinfo {author} {\bibfnamefont {U.}~\bibnamefont {Boscain}},
			\bibinfo {author} {\bibfnamefont {T.}~\bibnamefont {Calarco}}, \bibinfo
			{author} {\bibfnamefont {C.~P.}\ \bibnamefont {Koch}}, \bibinfo {author}
			{\bibfnamefont {W.}~\bibnamefont {K{\"o}ckenberger}}, \bibinfo {author}
			{\bibfnamefont {R.}~\bibnamefont {Kosloff}}, \bibinfo {author} {\bibfnamefont
				{I.}~\bibnamefont {Kuprov}}, \bibinfo {author} {\bibfnamefont
				{B.}~\bibnamefont {Luy}}, \bibinfo {author} {\bibfnamefont {S.}~\bibnamefont
				{Schirmer}}, \bibinfo {author} {\bibfnamefont {T.}~\bibnamefont
				{Schulte-Herbr{\"u}ggen}}, \bibinfo {author} {\bibfnamefont {D.}~\bibnamefont
				{Sugny}}, \ and\ \bibinfo {author} {\bibfnamefont {F.~K.}\ \bibnamefont
				{Wilhelm}},\ }\href {\doibase 10.1140/epjd/e2015-60464-1} {\bibfield
			{journal} {\bibinfo  {journal} {The European Physical Journal D}\ }\textbf
			{\bibinfo {volume} {69}},\ \bibinfo {pages} {279} (\bibinfo {year}
			{2015})}\BibitemShut {NoStop}%
		\bibitem [{\citenamefont {D'Alessandro}(2021)}]{Alessandro2021}%
		\BibitemOpen
		\bibfield  {author} {\bibinfo {author} {\bibfnamefont {D.}~\bibnamefont
				{D'Alessandro}},\ }\href {https://doi.org/10.1201/9781003051268} {\emph
			{\bibinfo {title} {Introduction to Quantum Control and Dynamics (2nd ed.)}}}\
		(\bibinfo  {publisher} {Chapman and Hall/CRC},\ \bibinfo {year}
		{2021})\BibitemShut {NoStop}%
		\bibitem [{\citenamefont {Krotov}(1993)}]{Krotov1993}%
		\BibitemOpen
		\bibfield  {author} {\bibinfo {author} {\bibfnamefont {V.~F.}\ \bibnamefont
				{Krotov}},\ }\enquote {\bibinfo {title} {Global methods in optimal control
				theory},}\ in\ \href {\doibase 10.1007/978-1-4612-0349-0_3} {\emph {\bibinfo
				{booktitle} {Advances in Nonlinear Dynamics and Control: A Report from
					Russia}}},\ \bibinfo {editor} {edited by\ \bibinfo {editor} {\bibfnamefont
				{A.~B.}\ \bibnamefont {Kurzhanski}}}\ (\bibinfo  {publisher} {Birkh{\"a}user
			Boston},\ \bibinfo {address} {Boston, MA},\ \bibinfo {year} {1993})\ pp.\
		\bibinfo {pages} {74--121}\BibitemShut {NoStop}%
		\bibitem [{\citenamefont {Brif}\ \emph {et~al.}(2010)\citenamefont {Brif},
			\citenamefont {Chakrabarti},\ and\ \citenamefont {Rabitz}}]{Brif2010}%
		\BibitemOpen
		\bibfield  {author} {\bibinfo {author} {\bibfnamefont {C.}~\bibnamefont
				{Brif}}, \bibinfo {author} {\bibfnamefont {R.}~\bibnamefont {Chakrabarti}}, \
			and\ \bibinfo {author} {\bibfnamefont {H.}~\bibnamefont {Rabitz}},\ }\href
		{\doibase 10.1088/1367-2630/12/7/075008} {\bibfield  {journal} {\bibinfo
				{journal} {New Journal of Physics}\ }\textbf {\bibinfo {volume} {12}},\
			\bibinfo {pages} {075008} (\bibinfo {year} {2010})}\BibitemShut {NoStop}%
		\bibitem [{\citenamefont {Gericke}\ \emph {et~al.}(2007)\citenamefont
			{Gericke}, \citenamefont {Gerbier}, \citenamefont {Widera}, \citenamefont
			{F?lling}, \citenamefont {Mandel},\ and\ \citenamefont
			{Bloch}}]{Gericke2007}%
		\BibitemOpen
		\bibfield  {author} {\bibinfo {author} {\bibfnamefont {T.}~\bibnamefont
				{Gericke}}, \bibinfo {author} {\bibfnamefont {F.}~\bibnamefont {Gerbier}},
			\bibinfo {author} {\bibfnamefont {A.}~\bibnamefont {Widera}}, \bibinfo
			{author} {\bibfnamefont {S.}~\bibnamefont {F{\"o}lling}}, \bibinfo {author}
			{\bibfnamefont {O.}~\bibnamefont {Mandel}}, \ and\ \bibinfo {author}
			{\bibfnamefont {I.}~\bibnamefont {Bloch}},\ }\href {\doibase
			10.1080/09500340600777730} {\bibfield  {journal} {\bibinfo  {journal}
				{Journal of Modern Optics}\ }\textbf {\bibinfo {volume} {54}},\ \bibinfo
			{pages} {735} (\bibinfo {year} {2007})},\ \Eprint
		{http://arxiv.org/abs/https://doi.org/10.1080/09500340600777730}
		{https://doi.org/10.1080/09500340600777730} \BibitemShut {NoStop}%
		\bibitem [{\citenamefont {Gu\'ery-Odelin}\ \emph {et~al.}(2019)\citenamefont
			{Gu\'ery-Odelin}, \citenamefont {Ruschhaupt}, \citenamefont {Kiely},
			\citenamefont {Torrontegui}, \citenamefont {Mart\'{\i}nez-Garaot},\ and\
			\citenamefont {Muga}}]{Odelin2019}%
		\BibitemOpen
		\bibfield  {author} {\bibinfo {author} {\bibfnamefont {D.}~\bibnamefont
				{Gu\'ery-Odelin}}, \bibinfo {author} {\bibfnamefont {A.}~\bibnamefont
				{Ruschhaupt}}, \bibinfo {author} {\bibfnamefont {A.}~\bibnamefont {Kiely}},
			\bibinfo {author} {\bibfnamefont {E.}~\bibnamefont {Torrontegui}}, \bibinfo
			{author} {\bibfnamefont {S.}~\bibnamefont {Mart\'{\i}nez-Garaot}}, \ and\
			\bibinfo {author} {\bibfnamefont {J.~G.}\ \bibnamefont {Muga}},\ }\href
		{\doibase 10.1103/RevModPhys.91.045001} {\bibfield  {journal} {\bibinfo
				{journal} {Rev. Mod. Phys.}\ }\textbf {\bibinfo {volume} {91}},\ \bibinfo
			{pages} {045001} (\bibinfo {year} {2019})}\BibitemShut {NoStop}%
		\bibitem [{\citenamefont {Chen}\ \emph {et~al.}(2010)\citenamefont {Chen},
			\citenamefont {Ruschhaupt}, \citenamefont {Schmidt}, \citenamefont {del
				Campo}, \citenamefont {Gu\'ery-Odelin},\ and\ \citenamefont
			{Muga}}]{Chen2010}%
		\BibitemOpen
		\bibfield  {author} {\bibinfo {author} {\bibfnamefont {X.}~\bibnamefont
				{Chen}}, \bibinfo {author} {\bibfnamefont {A.}~\bibnamefont {Ruschhaupt}},
			\bibinfo {author} {\bibfnamefont {S.}~\bibnamefont {Schmidt}}, \bibinfo
			{author} {\bibfnamefont {A.}~\bibnamefont {del Campo}}, \bibinfo {author}
			{\bibfnamefont {D.}~\bibnamefont {Gu\'ery-Odelin}}, \ and\ \bibinfo {author}
			{\bibfnamefont {J.~G.}\ \bibnamefont {Muga}},\ }\href {\doibase
			10.1103/PhysRevLett.104.063002} {\bibfield  {journal} {\bibinfo  {journal}
				{Phys. Rev. Lett.}\ }\textbf {\bibinfo {volume} {104}},\ \bibinfo {pages}
			{063002} (\bibinfo {year} {2010})}\BibitemShut {NoStop}%
		\bibitem [{\citenamefont {Khaneja}\ \emph {et~al.}(2005)\citenamefont
			{Khaneja}, \citenamefont {Reiss}, \citenamefont {Kehlet}, \citenamefont
			{Schulte-Herbr\"{u}ggen},\ and\ \citenamefont {Glaser}}]{KHANEJA2005}%
		\BibitemOpen
		\bibfield  {author} {\bibinfo {author} {\bibfnamefont {N.}~\bibnamefont
				{Khaneja}}, \bibinfo {author} {\bibfnamefont {T.}~\bibnamefont {Reiss}},
			\bibinfo {author} {\bibfnamefont {C.}~\bibnamefont {Kehlet}}, \bibinfo
			{author} {\bibfnamefont {T.}~\bibnamefont {Schulte-Herbr\"{u}ggen}}, \ and\
			\bibinfo {author} {\bibfnamefont {S.~J.}\ \bibnamefont {Glaser}},\ }\href
		{\doibase https://doi.org/10.1016/j.jmr.2004.11.004} {\bibfield  {journal}
			{\bibinfo  {journal} {Journal of Magnetic Resonance}\ }\textbf {\bibinfo
				{volume} {172}},\ \bibinfo {pages} {296} (\bibinfo {year}
			{2005})}\BibitemShut {NoStop}%
		\bibitem [{\citenamefont {van Frank}\ \emph {et~al.}(2016)\citenamefont {van
				Frank}, \citenamefont {Bonneau}, \citenamefont {Schmiedmayer}, \citenamefont
			{Hild}, \citenamefont {Gross}, \citenamefont {Cheneau}, \citenamefont
			{Bloch}, \citenamefont {Pichler}, \citenamefont {Negretti}, \citenamefont
			{Calarco},\ and\ \citenamefont {Montangero}}]{vanFrank2016}%
		\BibitemOpen
		\bibfield  {author} {\bibinfo {author} {\bibfnamefont {S.}~\bibnamefont {van
					Frank}}, \bibinfo {author} {\bibfnamefont {M.}~\bibnamefont {Bonneau}},
			\bibinfo {author} {\bibfnamefont {J.}~\bibnamefont {Schmiedmayer}}, \bibinfo
			{author} {\bibfnamefont {S.}~\bibnamefont {Hild}}, \bibinfo {author}
			{\bibfnamefont {C.}~\bibnamefont {Gross}}, \bibinfo {author} {\bibfnamefont
				{M.}~\bibnamefont {Cheneau}}, \bibinfo {author} {\bibfnamefont
				{I.}~\bibnamefont {Bloch}}, \bibinfo {author} {\bibfnamefont
				{T.}~\bibnamefont {Pichler}}, \bibinfo {author} {\bibfnamefont
				{A.}~\bibnamefont {Negretti}}, \bibinfo {author} {\bibfnamefont
				{T.}~\bibnamefont {Calarco}}, \ and\ \bibinfo {author} {\bibfnamefont
				{S.}~\bibnamefont {Montangero}},\ }\href {\doibase 10.1038/srep34187}
		{\bibfield  {journal} {\bibinfo  {journal} {Scientific Reports}\ }\textbf
			{\bibinfo {volume} {6}},\ \bibinfo {pages} {34187} (\bibinfo {year}
			{2016})}\BibitemShut {NoStop}%
		\bibitem [{\citenamefont {Li}\ \emph {et~al.}(2018)\citenamefont {Li},
			\citenamefont {Pecak}, \citenamefont {Sowi\'{n}ski}, \citenamefont
			{Sherson},\ and\ \citenamefont {Nielsen}}]{Li2018}%
		\BibitemOpen
		\bibfield  {author} {\bibinfo {author} {\bibfnamefont {X.}~\bibnamefont
				{Li}}, \bibinfo {author} {\bibfnamefont {D.}~\bibnamefont {Pecak}}, \bibinfo
			{author} {\bibfnamefont {T.}~\bibnamefont {Sowi\'{n}ski}}, \bibinfo {author}
			{\bibfnamefont {J.}~\bibnamefont {Sherson}}, \ and\ \bibinfo {author}
			{\bibfnamefont {A.~E.~B.}\ \bibnamefont {Nielsen}},\ }\href {\doibase
			10.1103/PhysRevA.97.033602} {\bibfield  {journal} {\bibinfo  {journal} {Phys.
					Rev. A}\ }\textbf {\bibinfo {volume} {97}},\ \bibinfo {pages} {033602}
			(\bibinfo {year} {2018})}\BibitemShut {NoStop}%
		\bibitem [{\citenamefont {Srivatsa}\ \emph {et~al.}(2021)\citenamefont
			{Srivatsa}, \citenamefont {Li},\ and\ \citenamefont
			{Nielsen}}]{Srivatsa2021}%
		\BibitemOpen
		\bibfield  {author} {\bibinfo {author} {\bibfnamefont {N.~S.}\ \bibnamefont
				{Srivatsa}}, \bibinfo {author} {\bibfnamefont {X.}~\bibnamefont {Li}}, \ and\
			\bibinfo {author} {\bibfnamefont {A.~E.~B.}\ \bibnamefont {Nielsen}},\ }\href
		{\doibase 10.1103/PhysRevResearch.3.033044} {\bibfield  {journal} {\bibinfo
				{journal} {Phys. Rev. Research}\ }\textbf {\bibinfo {volume} {3}},\ \bibinfo
			{pages} {033044} (\bibinfo {year} {2021})}\BibitemShut {NoStop}%
		\bibitem [{\citenamefont {Caneva}\ \emph {et~al.}(2009)\citenamefont {Caneva},
			\citenamefont {Murphy}, \citenamefont {Calarco}, \citenamefont {Fazio},
			\citenamefont {Montangero}, \citenamefont {Giovannetti},\ and\ \citenamefont
			{Santoro}}]{Caneva2009}%
		\BibitemOpen
		\bibfield  {author} {\bibinfo {author} {\bibfnamefont {T.}~\bibnamefont
				{Caneva}}, \bibinfo {author} {\bibfnamefont {M.}~\bibnamefont {Murphy}},
			\bibinfo {author} {\bibfnamefont {T.}~\bibnamefont {Calarco}}, \bibinfo
			{author} {\bibfnamefont {R.}~\bibnamefont {Fazio}}, \bibinfo {author}
			{\bibfnamefont {S.}~\bibnamefont {Montangero}}, \bibinfo {author}
			{\bibfnamefont {V.}~\bibnamefont {Giovannetti}}, \ and\ \bibinfo {author}
			{\bibfnamefont {G.~E.}\ \bibnamefont {Santoro}},\ }\href {\doibase
			10.1103/PhysRevLett.103.240501} {\bibfield  {journal} {\bibinfo  {journal}
				{Phys. Rev. Lett.}\ }\textbf {\bibinfo {volume} {103}},\ \bibinfo {pages}
			{240501} (\bibinfo {year} {2009})}\BibitemShut {NoStop}%
		\bibitem [{\citenamefont {Lloyd}\ and\ \citenamefont
			{Montangero}(2014)}]{Lloyd2014}%
		\BibitemOpen
		\bibfield  {author} {\bibinfo {author} {\bibfnamefont {S.}~\bibnamefont
				{Lloyd}}\ and\ \bibinfo {author} {\bibfnamefont {S.}~\bibnamefont
				{Montangero}},\ }\href {\doibase 10.1103/PhysRevLett.113.010502} {\bibfield
			{journal} {\bibinfo  {journal} {Phys. Rev. Lett.}\ }\textbf {\bibinfo
				{volume} {113}},\ \bibinfo {pages} {010502} (\bibinfo {year}
			{2014})}\BibitemShut {NoStop}%
		\bibitem [{\citenamefont {D'Alessandro}\ and\ \citenamefont
			{Dahleh}(2001)}]{Alessandro2001}%
		\BibitemOpen
		\bibfield  {author} {\bibinfo {author} {\bibfnamefont {D.}~\bibnamefont
				{D'Alessandro}}\ and\ \bibinfo {author} {\bibfnamefont {M.}~\bibnamefont
				{Dahleh}},\ }\href {\doibase 10.1109/9.928587} {\bibfield  {journal}
			{\bibinfo  {journal} {IEEE Transactions on Automatic Control}\ }\textbf
			{\bibinfo {volume} {46}},\ \bibinfo {pages} {866} (\bibinfo {year}
			{2001})}\BibitemShut {NoStop}%
		\bibitem [{\citenamefont {Boscain}\ \emph {et~al.}(2002)\citenamefont
			{Boscain}, \citenamefont {Charlot}, \citenamefont {Gauthier}, \citenamefont
			{Gu\'{e}rin},\ and\ \citenamefont {Jauslin}}]{Boscain2002}%
		\BibitemOpen
		\bibfield  {author} {\bibinfo {author} {\bibfnamefont {U.}~\bibnamefont
				{Boscain}}, \bibinfo {author} {\bibfnamefont {G.}~\bibnamefont {Charlot}},
			\bibinfo {author} {\bibfnamefont {J.-P.}\ \bibnamefont {Gauthier}}, \bibinfo
			{author} {\bibfnamefont {S.}~\bibnamefont {Gu\'{e}rin}}, \ and\ \bibinfo
			{author} {\bibfnamefont {H.-R.}\ \bibnamefont {Jauslin}},\ }\href {\doibase
			10.1063/1.1465516} {\bibfield  {journal} {\bibinfo  {journal} {Journal of
					Mathematical Physics}\ }\textbf {\bibinfo {volume} {43}},\ \bibinfo {pages}
			{2107} (\bibinfo {year} {2002})},\ \Eprint
		{http://arxiv.org/abs/https://aip.scitation.org/doi/pdf/10.1063/1.1465516}
		{https://aip.scitation.org/doi/pdf/10.1063/1.1465516} \BibitemShut {NoStop}%
		\bibitem [{\citenamefont {Khaneja}\ \emph {et~al.}(2001)\citenamefont
			{Khaneja}, \citenamefont {Brockett},\ and\ \citenamefont
			{Glaser}}]{Khaneja2001}%
		\BibitemOpen
		\bibfield  {author} {\bibinfo {author} {\bibfnamefont {N.}~\bibnamefont
				{Khaneja}}, \bibinfo {author} {\bibfnamefont {R.}~\bibnamefont {Brockett}}, \
			and\ \bibinfo {author} {\bibfnamefont {S.~J.}\ \bibnamefont {Glaser}},\
		}\href {\doibase 10.1103/PhysRevA.63.032308} {\bibfield  {journal} {\bibinfo
				{journal} {Phys. Rev. A}\ }\textbf {\bibinfo {volume} {63}},\ \bibinfo
			{pages} {032308} (\bibinfo {year} {2001})}\BibitemShut {NoStop}%
		\bibitem [{\citenamefont {Boscain}\ and\ \citenamefont
			{Chitour}(2005)}]{Boscain2005}%
		\BibitemOpen
		\bibfield  {author} {\bibinfo {author} {\bibfnamefont {U.}~\bibnamefont
				{Boscain}}\ and\ \bibinfo {author} {\bibfnamefont {Y.}~\bibnamefont
				{Chitour}},\ }\href {\doibase 10.1137/S0363012904441532} {\bibfield
			{journal} {\bibinfo  {journal} {SIAM Journal on Control and Optimization}\
			}\textbf {\bibinfo {volume} {44}},\ \bibinfo {pages} {111} (\bibinfo {year}
			{2005})},\ \Eprint
		{http://arxiv.org/abs/https://doi.org/10.1137/S0363012904441532}
		{https://doi.org/10.1137/S0363012904441532} \BibitemShut {NoStop}%
		\bibitem [{\citenamefont {Boscain}\ and\ \citenamefont
			{Mason}(2006)}]{Boscain2006}%
		\BibitemOpen
		\bibfield  {author} {\bibinfo {author} {\bibfnamefont {U.}~\bibnamefont
				{Boscain}}\ and\ \bibinfo {author} {\bibfnamefont {P.}~\bibnamefont
				{Mason}},\ }\href {\doibase 10.1063/1.2203236} {\bibfield  {journal}
			{\bibinfo  {journal} {Journal of Mathematical Physics}\ }\textbf {\bibinfo
				{volume} {47}},\ \bibinfo {pages} {062101} (\bibinfo {year} {2006})},\
		\Eprint {http://arxiv.org/abs/https://doi.org/10.1063/1.2203236}
		{https://doi.org/10.1063/1.2203236} \BibitemShut {NoStop}%
		\bibitem [{\citenamefont {Boscain}\ \emph {et~al.}(2014)\citenamefont
			{Boscain}, \citenamefont {Gr?nberg}, \citenamefont {Long},\ and\
			\citenamefont {Rabitz}}]{Boscain2014}%
		\BibitemOpen
		\bibfield  {author} {\bibinfo {author} {\bibfnamefont {U.}~\bibnamefont
				{Boscain}}, \bibinfo {author} {\bibfnamefont {F.}~\bibnamefont {Gr{\"o}nberg}},
			\bibinfo {author} {\bibfnamefont {R.}~\bibnamefont {Long}}, \ and\ \bibinfo
			{author} {\bibfnamefont {H.}~\bibnamefont {Rabitz}},\ }\href {\doibase
			10.1063/1.4882158} {\bibfield  {journal} {\bibinfo  {journal} {Journal of
					Mathematical Physics}\ }\textbf {\bibinfo {volume} {55}},\ \bibinfo {pages}
			{062106} (\bibinfo {year} {2014})},\ \Eprint
		{http://arxiv.org/abs/https://doi.org/10.1063/1.4882158}
		{https://doi.org/10.1063/1.4882158} \BibitemShut {NoStop}%
		\bibitem [{\citenamefont {Hegerfeldt}(2013)}]{Hegerfeldt2013}%
		\BibitemOpen
		\bibfield  {author} {\bibinfo {author} {\bibfnamefont {G.~C.}\ \bibnamefont
				{Hegerfeldt}},\ }\href {\doibase 10.1103/PhysRevLett.111.260501} {\bibfield
			{journal} {\bibinfo  {journal} {Phys. Rev. Lett.}\ }\textbf {\bibinfo
				{volume} {111}},\ \bibinfo {pages} {260501} (\bibinfo {year}
			{2013})}\BibitemShut {NoStop}%
		\bibitem [{\citenamefont {Hegerfeldt}(2014)}]{Hegerfeldt2014}%
		\BibitemOpen
		\bibfield  {author} {\bibinfo {author} {\bibfnamefont {G.~C.}\ \bibnamefont
				{Hegerfeldt}},\ }\href {\doibase 10.1103/PhysRevA.90.032110} {\bibfield
			{journal} {\bibinfo  {journal} {Phys. Rev. A}\ }\textbf {\bibinfo {volume}
				{90}},\ \bibinfo {pages} {032110} (\bibinfo {year} {2014})}\BibitemShut
		{NoStop}%
		\bibitem [{\citenamefont {Boozer}(2012)}]{Boozer2012}%
		\BibitemOpen
		\bibfield  {author} {\bibinfo {author} {\bibfnamefont {A.~D.}\ \bibnamefont
				{Boozer}},\ }\href {\doibase 10.1103/PhysRevA.85.012317} {\bibfield
			{journal} {\bibinfo  {journal} {Phys. Rev. A}\ }\textbf {\bibinfo {volume}
				{85}},\ \bibinfo {pages} {012317} (\bibinfo {year} {2012})}\BibitemShut
		{NoStop}%
		\bibitem [{\citenamefont {Jafarizadeh}\ \emph {et~al.}(2020)\citenamefont
			{Jafarizadeh}, \citenamefont {Naghdi},\ and\ \citenamefont
			{Bazrafkan}}]{Jafarizadeh2020}%
		\BibitemOpen
		\bibfield  {author} {\bibinfo {author} {\bibfnamefont {M.}~\bibnamefont
				{Jafarizadeh}}, \bibinfo {author} {\bibfnamefont {F.}~\bibnamefont {Naghdi}},
			\ and\ \bibinfo {author} {\bibfnamefont {M.}~\bibnamefont {Bazrafkan}},\
		}\href {\doibase https://doi.org/10.1016/j.physleta.2020.126743} {\bibfield
			{journal} {\bibinfo  {journal} {Physics Letters A}\ }\textbf {\bibinfo
				{volume} {384}},\ \bibinfo {pages} {126743} (\bibinfo {year}
			{2020})}\BibitemShut {NoStop}%
		\bibitem [{\citenamefont {Sklarz}\ and\ \citenamefont
			{Tannor}(2002)}]{Sklarz2002}%
		\BibitemOpen
		\bibfield  {author} {\bibinfo {author} {\bibfnamefont {S.~E.}\ \bibnamefont
				{Sklarz}}\ and\ \bibinfo {author} {\bibfnamefont {D.~J.}\ \bibnamefont
				{Tannor}},\ }\href {\doibase 10.1103/PhysRevA.66.053619} {\bibfield
			{journal} {\bibinfo  {journal} {Phys. Rev. A}\ }\textbf {\bibinfo {volume}
				{66}},\ \bibinfo {pages} {053619} (\bibinfo {year} {2002})}\BibitemShut
		{NoStop}%
		\bibitem [{\citenamefont {Doria}\ \emph {et~al.}(2011)\citenamefont {Doria},
			\citenamefont {Calarco},\ and\ \citenamefont {Montangero}}]{Doria2011}%
		\BibitemOpen
		\bibfield  {author} {\bibinfo {author} {\bibfnamefont {P.}~\bibnamefont
				{Doria}}, \bibinfo {author} {\bibfnamefont {T.}~\bibnamefont {Calarco}}, \
			and\ \bibinfo {author} {\bibfnamefont {S.}~\bibnamefont {Montangero}},\
		}\href {\doibase 10.1103/PhysRevLett.106.190501} {\bibfield  {journal}
			{\bibinfo  {journal} {Phys. Rev. Lett.}\ }\textbf {\bibinfo {volume} {106}},\
			\bibinfo {pages} {190501} (\bibinfo {year} {2011})}\BibitemShut {NoStop}%
		\bibitem [{\citenamefont {S{\o}rensen}\ \emph {et~al.}(2018)\citenamefont
			{S\o{}rensen}, \citenamefont {Aranburu}, \citenamefont {Heinzel},\ and\
			\citenamefont {Sherson}}]{Sorensen2018}%
		\BibitemOpen
		\bibfield  {author} {\bibinfo {author} {\bibfnamefont {J.~J. W.~H.}\
				\bibnamefont {S{\o}rensen}}, \bibinfo {author} {\bibfnamefont {M.~O.}\
				\bibnamefont {Aranburu}}, \bibinfo {author} {\bibfnamefont {T.}~\bibnamefont
				{Heinzel}}, \ and\ \bibinfo {author} {\bibfnamefont {J.~F.}\ \bibnamefont
				{Sherson}},\ }\href {\doibase 10.1103/PhysRevA.98.022119} {\bibfield
			{journal} {\bibinfo  {journal} {Phys. Rev. A}\ }\textbf {\bibinfo {volume}
				{98}},\ \bibinfo {pages} {022119} (\bibinfo {year} {2018})}\BibitemShut
		{NoStop}%
		\bibitem [{\citenamefont {Machnes}\ \emph {et~al.}(2018)\citenamefont
			{Machnes}, \citenamefont {Ass\'emat}, \citenamefont {Tannor},\ and\
			\citenamefont {Wilhelm}}]{Machnes2018}%
		\BibitemOpen
		\bibfield  {author} {\bibinfo {author} {\bibfnamefont {S.}~\bibnamefont
				{Machnes}}, \bibinfo {author} {\bibfnamefont {E.}~\bibnamefont {Ass\'emat}},
			\bibinfo {author} {\bibfnamefont {D.}~\bibnamefont {Tannor}}, \ and\ \bibinfo
			{author} {\bibfnamefont {F.~K.}\ \bibnamefont {Wilhelm}},\ }\href {\doibase
			10.1103/PhysRevLett.120.150401} {\bibfield  {journal} {\bibinfo  {journal}
				{Phys. Rev. Lett.}\ }\textbf {\bibinfo {volume} {120}},\ \bibinfo {pages}
			{150401} (\bibinfo {year} {2018})}\BibitemShut {NoStop}%
		\bibitem [{\citenamefont {Zahedinejad}\ \emph {et~al.}(2014)\citenamefont
			{Zahedinejad}, \citenamefont {Schirmer},\ and\ \citenamefont
			{Sanders}}]{Zahedinejad2014}%
		\BibitemOpen
		\bibfield  {author} {\bibinfo {author} {\bibfnamefont {E.}~\bibnamefont
				{Zahedinejad}}, \bibinfo {author} {\bibfnamefont {S.}~\bibnamefont
				{Schirmer}}, \ and\ \bibinfo {author} {\bibfnamefont {B.~C.}\ \bibnamefont
				{Sanders}},\ }\href {\doibase 10.1103/PhysRevA.90.032310} {\bibfield
			{journal} {\bibinfo  {journal} {Phys. Rev. A}\ }\textbf {\bibinfo {volume}
				{90}},\ \bibinfo {pages} {032310} (\bibinfo {year} {2014})}\BibitemShut
		{NoStop}%
		\bibitem [{\citenamefont {Bukov}\ \emph
			{et~al.}(2018{\natexlab{a}})\citenamefont {Bukov}, \citenamefont {Day},
			\citenamefont {Sels}, \citenamefont {Weinberg}, \citenamefont {Polkovnikov},\
			and\ \citenamefont {Mehta}}]{Bukov2018a}%
		\BibitemOpen
		\bibfield  {author} {\bibinfo {author} {\bibfnamefont {M.}~\bibnamefont
				{Bukov}}, \bibinfo {author} {\bibfnamefont {A.~G.~R.}\ \bibnamefont {Day}},
			\bibinfo {author} {\bibfnamefont {D.}~\bibnamefont {Sels}}, \bibinfo {author}
			{\bibfnamefont {P.}~\bibnamefont {Weinberg}}, \bibinfo {author}
			{\bibfnamefont {A.}~\bibnamefont {Polkovnikov}}, \ and\ \bibinfo {author}
			{\bibfnamefont {P.}~\bibnamefont {Mehta}},\ }\href {\doibase
			10.1103/PhysRevX.8.031086} {\bibfield  {journal} {\bibinfo  {journal} {Phys.
					Rev. X}\ }\textbf {\bibinfo {volume} {8}},\ \bibinfo {pages} {031086}
			(\bibinfo {year} {2018}{\natexlab{a}})}\BibitemShut {NoStop}%
		\bibitem [{\citenamefont {Pechen}\ and\ \citenamefont
			{Tannor}(2011)}]{Pechen2011}%
		\BibitemOpen
		\bibfield  {author} {\bibinfo {author} {\bibfnamefont {A.~N.}\ \bibnamefont
				{Pechen}}\ and\ \bibinfo {author} {\bibfnamefont {D.~J.}\ \bibnamefont
				{Tannor}},\ }\href {\doibase 10.1103/PhysRevLett.106.120402} {\bibfield
			{journal} {\bibinfo  {journal} {Phys. Rev. Lett.}\ }\textbf {\bibinfo
				{volume} {106}},\ \bibinfo {pages} {120402} (\bibinfo {year}
			{2011})}\BibitemShut {NoStop}%
		\bibitem [{\citenamefont {Larocca}\ \emph {et~al.}(2018)\citenamefont
			{Larocca}, \citenamefont {Poggi},\ and\ \citenamefont
			{Wisniacki}}]{Larocca2018}%
		\BibitemOpen
		\bibfield  {author} {\bibinfo {author} {\bibfnamefont {M.}~\bibnamefont
				{Larocca}}, \bibinfo {author} {\bibfnamefont {P.~M.}\ \bibnamefont {Poggi}},
			\ and\ \bibinfo {author} {\bibfnamefont {D.~A.}\ \bibnamefont {Wisniacki}},\
		}\href {\doibase 10.1088/1751-8121/aad657} {\bibfield  {journal} {\bibinfo
				{journal} {Journal of Physics A: Mathematical and Theoretical}\ }\textbf
			{\bibinfo {volume} {51}},\ \bibinfo {pages} {385305} (\bibinfo {year}
			{2018})}\BibitemShut {NoStop}%
		\bibitem [{\citenamefont {Jafarizadeh}\ \emph {et~al.}(2022)\citenamefont
			{Jafarizadeh}, \citenamefont {Naghdi},\ and\ \citenamefont
			{Bazrafkan}}]{Jafarizadeh2022}%
		\BibitemOpen
		\bibfield  {author} {\bibinfo {author} {\bibfnamefont {M.~A.}\ \bibnamefont
				{Jafarizadeh}}, \bibinfo {author} {\bibfnamefont {F.}~\bibnamefont {Naghdi}},
			\ and\ \bibinfo {author} {\bibfnamefont {M.~R.}\ \bibnamefont {Bazrafkan}},\
		}\href {\doibase 10.1140/epjp/s13360-022-02904-3} {\bibfield  {journal}
			{\bibinfo  {journal} {The European Physical Journal Plus}\ }\textbf {\bibinfo
				{volume} {137}},\ \bibinfo {pages} {720} (\bibinfo {year}
			{2022})}\BibitemShut {NoStop}%
		\bibitem [{\citenamefont {Li}(2022)}]{Li2022}%
		\BibitemOpen
		\bibfield  {author} {\bibinfo {author} {\bibfnamefont {X.}~\bibnamefont
				{Li}},\ }\href@noop {} {\bibfield  {journal} {\bibinfo  {journal}
				{arXiv:2208.13377}\ } (\bibinfo {year} {2022})}\BibitemShut {NoStop}%
		\bibitem [{\citenamefont {Bukov}\ \emph
			{et~al.}(2018{\natexlab{b}})\citenamefont {Bukov}, \citenamefont {Day},
			\citenamefont {Weinberg}, \citenamefont {Polkovnikov}, \citenamefont
			{Mehta},\ and\ \citenamefont {Sels}}]{Bukov2018b}%
		\BibitemOpen
		\bibfield  {author} {\bibinfo {author} {\bibfnamefont {M.}~\bibnamefont
				{Bukov}}, \bibinfo {author} {\bibfnamefont {A.~G.~R.}\ \bibnamefont {Day}},
			\bibinfo {author} {\bibfnamefont {P.}~\bibnamefont {Weinberg}}, \bibinfo
			{author} {\bibfnamefont {A.}~\bibnamefont {Polkovnikov}}, \bibinfo {author}
			{\bibfnamefont {P.}~\bibnamefont {Mehta}}, \ and\ \bibinfo {author}
			{\bibfnamefont {D.}~\bibnamefont {Sels}},\ }\href {\doibase
			10.1103/PhysRevA.97.052114} {\bibfield  {journal} {\bibinfo  {journal} {Phys.
					Rev. A}\ }\textbf {\bibinfo {volume} {97}},\ \bibinfo {pages} {052114}
			(\bibinfo {year} {2018}{\natexlab{b}})}\BibitemShut {NoStop}%
		\bibitem [{\citenamefont {Hill}\ and\ \citenamefont
			{Wootters}(1997)}]{Hill1997}%
		\BibitemOpen
		\bibfield  {author} {\bibinfo {author} {\bibfnamefont {S.~A.}\ \bibnamefont
				{Hill}}\ and\ \bibinfo {author} {\bibfnamefont {W.~K.}\ \bibnamefont
				{Wootters}},\ }\href {\doibase 10.1103/PhysRevLett.78.5022} {\bibfield
			{journal} {\bibinfo  {journal} {Phys. Rev. Lett.}\ }\textbf {\bibinfo
				{volume} {78}},\ \bibinfo {pages} {5022} (\bibinfo {year}
			{1997})}\BibitemShut {NoStop}%
		\bibitem [{\citenamefont {Bao}\ \emph {et~al.}(2018)\citenamefont {Bao},
			\citenamefont {Kleer}, \citenamefont {Wang},\ and\ \citenamefont
			{Rahmani}}]{Bao2018}%
		\BibitemOpen
		\bibfield  {author} {\bibinfo {author} {\bibfnamefont {S.}~\bibnamefont
				{Bao}}, \bibinfo {author} {\bibfnamefont {S.}~\bibnamefont {Kleer}}, \bibinfo
			{author} {\bibfnamefont {R.}~\bibnamefont {Wang}}, \ and\ \bibinfo {author}
			{\bibfnamefont {A.}~\bibnamefont {Rahmani}},\ }\href {\doibase
			10.1103/PhysRevA.97.062343} {\bibfield  {journal} {\bibinfo  {journal} {Phys.
					Rev. A}\ }\textbf {\bibinfo {volume} {97}},\ \bibinfo {pages} {062343}
			(\bibinfo {year} {2018})}\BibitemShut {NoStop}%
		\bibitem [{\citenamefont {Cong}(2014)}]{Cong2014}%
		\BibitemOpen
		\bibfield  {author} {\bibinfo {author} {\bibfnamefont {S.}~\bibnamefont
				{Cong}},\ }\href {https://doi.org/10.1002/9781118608135} {\emph {\bibinfo
				{title} {Control of Quantum Systems}}}\ (\bibinfo  {publisher} {John Wiley \&
			Sons, Ltd},\ \bibinfo {year} {2014})\BibitemShut {NoStop}%
	\end{thebibliography}
\end{document}